\AddToHook{package/fontspec/after}{%
  \defaultfontfeatures[inconsolata]{%
    UprightFont=Inconsolatazi4-Regular.otf,
    BoldFont=Inconsolatazi4-Bold.otf,
    ItalicFont=Inconsolatazi4-Regular.otf,
    ItalicFeatures={FakeSlant=0.2},
    BoldItalicFont=Inconsolatazi4-Bold.otf,
    BoldItalicFeatures={FakeSlant=0.2}}}

\newif\ifanonymous

\anonymousfalse

\ifanonymous
\documentclass[acmsmall,screen,review,anonymous,nonacm]{acmart}
\else
\documentclass[acmsmall,screen,nonacm]{acmart}
\fi
\AtBeginDocument{%
  }

\setcopyright{none}

\usepackage[disable]{todonotes}
\usepackage{xspace}
\usepackage{subcaption}
\usepackage{csquotes}
\usepackage{hyperref}
\usepackage[inline]{enumitem}

\usepackage[capitalise]{cleveref}
\crefname{figure}{Fig.}{Figs.}
\Crefname{appendix}{Appx.}{Appx.}

\crefdefaultlabelformat{#2\textup{#1}#3}

\makeatletter
\AddToHook{cmd/appendix/before}{\def\cref@section@alias{appendix}\def\cref@subsection@alias{appendix}}
\makeatother

\usepackage{amsthm}
\newtheorem{theorem}{Theorem}

\newtheoremstyle{rmkstyle}{}{}{}{}{\bfseries}{.}{ }{\thmname{#1}\thmnote{ (#3)}}
\theoremstyle{rmkstyle}

\newtheorem{remark}{Remark}

\usepackage{unicode-math}

\usepackage{newunicodechar}
\newunicodechar{∀}{\ensuremath{\forall}}
\newunicodechar{∃}{\ensuremath{\exists}}
\newunicodechar{λ}{\ensuremath{\lambda}}
\newunicodechar{ρ}{\ensuremath{\rho}}
\newunicodechar{ω}{\ensuremath{\omega}}
\newunicodechar{χ}{\ensuremath{\chi}}
\newunicodechar{τ}{\ensuremath{\tau}}
\newunicodechar{→}{\ensuremath{\to}}
\newunicodechar{←}{\ensuremath{\leftarrow}}
\newunicodechar{↔}{\ensuremath{\leftrightarrow}}
\newunicodechar{⇒}{\ensuremath{\Rightarrow}}
\newunicodechar{≤}{\ensuremath{\le}}
\newunicodechar{≥}{\ensuremath{\ge}}
\newunicodechar{≠}{\ensuremath{\ne}}
\newunicodechar{⊥}{\ensuremath{\bot}}
\newunicodechar{⊤}{\ensuremath{\top}}
\newunicodechar{∧}{\ensuremath{\land}}
\newunicodechar{∨}{\ensuremath{\lor}}
\newunicodechar{¬}{\ensuremath{\neg}}
\newunicodechar{⊎}{\ensuremath{\uplus}}
\newunicodechar{≈}{\ensuremath{\approx}}
\newunicodechar{≅}{\ensuremath{\cong}}
\newunicodechar{⊨}{\ensuremath{\vDash}}
\newunicodechar{⋆}{\ensuremath{\star}}
\newunicodechar{✦}{\ensuremath{\finstar}}
\newunicodechar{⊛}{\ensuremath{\infstar}}
\newunicodechar{⊑}{\ensuremath{\sqsubseteq}}
\newunicodechar{≲}{\ensuremath{\lesssim}}
\newunicodechar{χ}{\ensuremath{\chi}}

\usepackage{minted}

\usepackage{mathpartir}
\usepackage[most]{tcolorbox}

\usepackage{tikz}
\usetikzlibrary{
  arrows.meta,positioning,shapes.geometric,automata,
  fit,backgrounds,
  calc,
  decorations.pathmorphing
}
\tcbuselibrary{skins}

\newcommand{\coqtt}[1]{\mintinline{coq}|#1|}
\newcommand{\coqett}[1]{\mintinline[escapeinside=@@]{coq}|#1|}
\newcommand{\itree}{\coqtt{itree}}
\newcommand{\ctree}{\coqtt{ctree}}
\newcommand{\itrace}{\coqtt{itrace}}
\newcommand{\ibuchi}{\coqtt{iBuchi}}

\newcommand{\rocq}{Rocq Prover}
\newcommand{\artifacturl}{\url{https://anonymous.4open.science/r/itree-future/}}

\newcommand{\define}{\ensuremath{\triangleq}}
\newcommand{\ddefine}{\ensuremath{\quad\define\quad}}
\newcommand{\qand}{\ensuremath{\quad\textnormal{and}\quad}}

\newcommand{\Ret}{\ensuremath{\text{\coqtt{Ret}}}}
\newcommand{\Vis}{\ensuremath{\text{\coqtt{Vis}}}}
\newcommand{\iTau}{\ensuremath{\text{\coqtt{Tau}}}}
\newcommand{\bind}{\mathbin{>\!\!>\!\!=}}

\newcommand{\trigger}{\ensuremath{\text{\coqtt{trigger}}}}
\newcommand{\iter}{\ensuremath{\text{\coqtt{iter}}}}

\newcommand{\nil}{\ensuremath{\text{\coqtt{nil}}}}
\newcommand{\any}{\ensuremath{\text{\coqtt{any}}}}
\newcommand{\app}{\mathbin{+\!\!+}}
\newcommand{\uni}{\mathbin{\uplus}}
\newcommand{\prd}{\mathbin{\&}}
\newcommand{\eve}{\ensuremath{\text{\coqtt{eve}}}}
\newcommand{\body}{\ensuremath{\textit{body}}}

\newcommand{\mostar}{\star}
\newcommand{\infstar}{\circledast}
\newcommand{\finstar}{\ast}

\newcommand{\embed}{\ensuremath{\text{\coqtt{embed}}}}
\newcommand{\rep}{\ensuremath{\text{\coqtt{rep}}}}

\newcommand{\trSat}{\mathbin{[\textup{tr}\!\vDash]}}
\newcommand{\sat}{\vDash}

\newcommand{\ibtinc}{\mathbin{[\sqsubseteq_\text{tr}]}}
\newcommand{\ibtinceq}{\mathbin{[\equiv_\text{tr}]}}
\newcommand{\ibsim}{\mathbin{[\lesssim]}}
\newcommand{\ibsimchr}{\mathbin{[\lesssim_\chi]}}
\newcommand{\ibsimeq}{\mathbin{[\equiv]}}
\newcommand{\ibsimchreq}{\mathbin{[\equiv_\chi]}}
\newcommand{\ibbisim}{\mathbin{[\sim]}}
\newcommand{\ibbisimchr}{\mathbin{[\sim_\chi]}}

\def\triangleforqed{\hbox{$\lhd$}}
\makeatletter
\DeclareRobustCommand{\qedT}{%
	\ifmmode
	\eqno \def\@badmath{$$}
	\let\eqno\relax \let\leqno\relax \let\veqno\relax
	\hbox{\triangleforqed}%
	\else
	\leavevmode\unskip\penalty9999 \hbox{}\nobreak\hfill
	\quad\hbox{\triangleforqed}%
	\fi
}
\makeatother

\begin{document}

\title{Compositional Temporal Specifications for Interaction Trees}


\author{Linyu Yang}
\orcid{0009-0007-3838-0538}
\affiliation{%
  \department{College of Computer Science and Technology}
  \institution{Zhejiang University}
  \city{Hangzhou}
  \country{China}}
\email{linyu.yang@zju.edu.cn}

\author{Jianwei Yin}
\orcid{0000-0003-4703-7348}
\affiliation{%
  \department{College of Computer Science and Technology}
  \institution{Zhejiang University}
  \city{Hangzhou}
  \country{China}}
\email{zjuyjw@zju.edu.cn}

\author{Mingshuai Chen}
\authornote{Corresponding author.}
\orcid{0000-0001-9663-7441}
\affiliation{%
  \department{College of Computer Science and Technology}
  \institution{Zhejiang University}
  \city{Hangzhou}
  \country{China}}
\email{m.chen@zju.edu.cn}
\renewcommand{\shortauthors}{Yang et al.}

\renewcommand{\shortauthors}{Yang et al.}



\begin{abstract}
The correctness of modern effectful programs depends not only on their final results but also on the temporal evolution of observable interactions with their environments. Interaction Trees (ITrees) provide a reusable denotational semantics for such programs by representing effectful computations as potentially infinite trees of observable interactions. However, reusable temporal reasoning over ITrees remains largely unexplored. This paper presents a compositional temporal specification framework for ITrees. We introduce \ibuchi{}, a shallowly embedded temporal specification language based on Choice Trees, which represents finite and infinite interaction traces and supports expressive temporal operators, including sequencing, branching, conjunction, repetition, and Linear Temporal Logic modalities. We define trace- and tree-level satisfaction relations between \ibuchi{} specifications and ITrees, together with compositional reasoning principles for modular verification of effectful programs. We integrate \ibuchi{} into the Dijkstra Monads Forever framework, thereby enabling temporal specifications to compose with conventional Hoare-style specifications and supporting a generator-based encoding of future conditions. We have formally developed and verified our framework in Rocq and demonstrated its expressiveness and compositionality through several non-trivial case studies.
\end{abstract}



\setlength{\floatsep}{1.3\baselineskip}
\setlength{\textfloatsep}{1\baselineskip}
\setlength{\intextsep}{1\baselineskip}

\maketitle

\section{Introduction}

Modern software systems are increasingly interactive, concurrent, and effectful. Their correctness depends not only on final results, but also on how computations evolve over time through \emph{observable interactions} with their environments. Examples include communication protocols, operating systems, distributed services, and resource management systems, in which important correctness properties are inherently \emph{temporal}: requests should eventually receive responses, resources should eventually be released, and undesirable behaviors should never occur.

Temporal logics provide a natural formalism for specifying such behavioral properties and have long been a cornerstone of formal verification \cite{clarkeHandbookModelChecking2018a,clarkeModelChecking1997}. However, temporal reasoning for \emph{effectful programs} -- which interact with external environments via, e.g., system calls, I/O operations, or communication with external services -- remains challenging. Existing approaches are often tightly coupled with a specific application (e.g., resource usage protocols \cite{marriottResourceUsageVerification2003} and IO operations \cite{malechaTracebasedVerificationImperative2011}), program logic (e.g., Hoare-style logic with program states \cite{songSpecifyingVerifyingFuture2026,songProveNFixTemporalPropertyGuided2024}), or effect system (e.g., \emph{ContEffs} \cite{songAutomatedTemporalVerification2022}),
thus rendering temporal specifications difficult to reuse across different verification frameworks. As a result, temporal properties are frequently treated as specialized extensions of existing logics rather than as independent logical components that can compose with different semantic foundations.

\emph{Interaction Trees} (ITrees) \cite{xiaInteractionTreesRepresenting2019} provide a promising substrate for addressing this problem. ITrees are a compositional semantic framework for effectful computations that represent programs as potentially infinite trees of observable interactions. They have been used to give executable and mechanized semantics for a variety of programming languages and effectful systems \cite{zhangVerifyingHTTPKeyValue2021,zakowskiModularCompositionalExecutable2021,chappeMonadicInterpretersConcurrent2025}. By separating program structure from effect interpretation, ITrees provide a natural abstraction boundary: internal effects can be interpreted away, while externally visible interactions remain available for reasoning. Nevertheless, the ITree ecosystem currently lacks a \emph{general temporal specification framework} that can express and verify rich temporal properties of these observable behaviors.

\begin{figure}[t]
\centering
\resizebox{.8\linewidth}{!}{
\begin{tikzpicture}[
  >=Latex,
  font=\small,
  node distance=1.25cm and 1.65cm,
  box/.style={
    draw,
    rounded corners=2pt,
    minimum height=1.0cm,
    minimum width=2.7cm,
    align=center,
    fill=gray!8
  },
  work/.style={
    draw=blue!55!black,
    rounded corners=2pt,
    minimum height=1.0cm,
    minimum width=2.7cm,
    align=center,
    fill=blue!10
  },
  spec/.style={
    draw=orange!70!black,
    rounded corners=2pt,
    minimum height=1.0cm,
    minimum width=2.7cm,
    align=center,
    fill=orange!10
  },
  framework/.style={
    draw=purple!60!black,
    rounded corners=2pt,
    minimum height=1.0cm,
    minimum width=3.2cm,
    align=center,
    fill=purple!8
  },
  arrow/.style={->, thick, draw=gray!70},
  dashedarrow/.style={->, thick, dashed}
]

\node[box] (prog) {Effectful Programs};
\node[box, below=of prog] (itree) {Interaction Trees\\ \cite{xiaInteractionTreesRepresenting2019}};
\node[framework, below=of itree] (dmf) {Dijkstra Monads Forever (DMF)\\ \cite{silverDijkstraMonadsForever2021}};

\node[box, right=of prog] (ctree) {Choice Trees\\ \cite{chappeChoiceTreesRepresenting2023, chappeChoiceTreesRepresenting2025a}};
\node[work, below=of ctree] (ibuchi) {\ibuchi{} Logic Type};
\node[spec, below=of ibuchi] (fc) {Future Conditions};

\node[spec, right=of ibuchi] (temp) {Compositional\\ Temporal Spec.};

\draw[arrow] (prog) -- node[left] {denoted by} (itree);
\draw[arrow] (dmf) -- node[left] {specifies} (itree);
\draw[arrow] (ibuchi) -- node[above] {specifies} (itree);
\draw[arrow] (temp) -- node[below,align=center,yshift=13pt] {represented\\by} (ibuchi);

\draw[arrow] (ctree) -- node[right] {representation base} (ibuchi);
\draw[arrow] (ibuchi) -- node[left,xshift=20pt] {direct embedding} (dmf);
\draw[arrow] (ibuchi) -- node[right] {generates} (fc);
\draw[arrow] (fc) -- node[below,align=center,yshift=14pt] {encode\\into} (dmf);
\draw[arrow] (temp) -- node[right,xshift=5pt] {describes} (fc);

\node[
  draw,
  dashed,
  rounded corners=4pt,
  fit=(ibuchi) (temp) (fc),
  inner sep=8pt,
  label={[font=\small]below:\textit{\large{This paper: temporal specifications as reusable component}}}
] {};






\end{tikzpicture}
}
\caption{Overview of our compositional temporal specification framework. Effectful programs are translated into ITrees, whose temporal properties are represented compositionally by the \ibuchi{} type, which also encodes generator-based future conditions. All these specifications are embedded into the DMF framework.}
\label{fig:overview}
\end{figure}
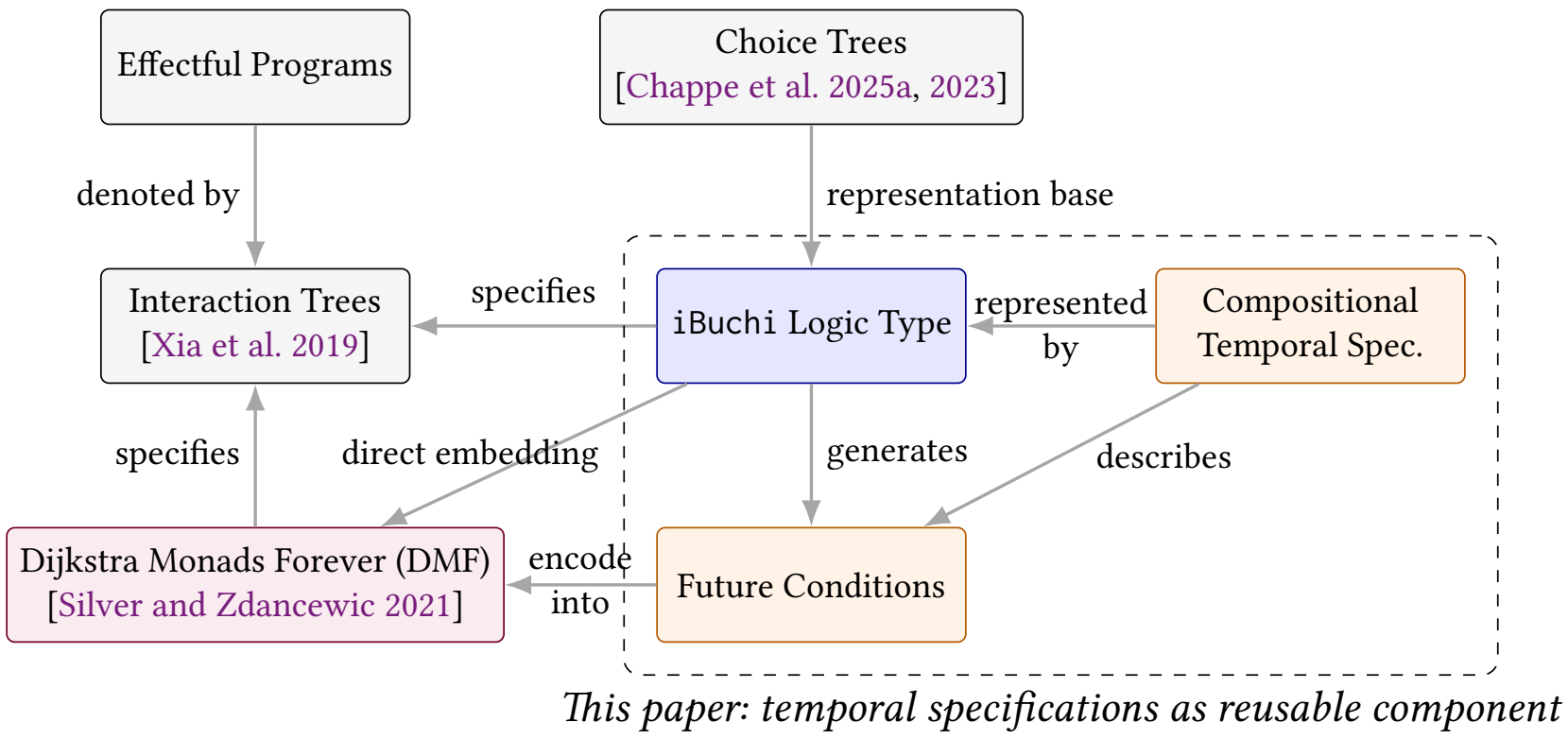

In this paper, we present \emph{a compositional temporal specification framework for interaction trees}. Our central idea is to treat temporal reasoning as a reusable logical layer over compositional program semantics, rather than tying it to a particular language or verification framework. Specifically, as sketched in \Cref{fig:overview}, we introduce the \ibuchi{} type, a temporal specification language shallowly embedded into \emph{Choice Trees} (CTrees) \cite{chappeChoiceTreesRepresenting2023,chappeChoiceTreesRepresenting2025a}. \ibuchi{} represents sets of finite and infinite interaction traces and supports expressive temporal specifications through compositional operators for event matching, sequencing, branching, conjunction, repetition, and Linear Temporal Logic (LTL) modalities \cite{clarkeHandbookModelChecking2018a}.

We define the semantics of \ibuchi{} through two levels of satisfaction. First, we establish a trace-level satisfaction relation between concrete interaction traces and temporal specifications. We then lift this relation to ITrees by requiring all observable traces of a computation to satisfy the specification. Based on these semantics, we develop \emph{compositional reasoning principles} for common ITree operations, including monadic composition, iteration, branching, and specification refinement. These principles allow temporal properties to be verified in a \emph{modular} manner without repeatedly reasoning about low-level trace behavior.

To integrate temporal reasoning with existing verification frameworks, we encode \ibuchi{} specifications into the \emph{Dijkstra Monads Forever} (DMF) framework \cite{silverDijkstraMonadsForever2021}. This integration enables temporal properties to coexist with conventional Hoare-style reasoning \cite{hoareAxiomaticBasisComputer1969,dijkstraGuardedCommandsNondeterminacy1975} for interactive computations. As an application, we show that generator-based \emph{future conditions} \cite{songProveNFixTemporalPropertyGuided2024, songSpecifyingVerifyingFuture2026} -- which specify necessary conditions after certain event happens -- can be naturally represented as temporal specifications over interaction events.

We evaluate our framework through several case studies with non-trivial specifications, including termination, memory safety, and mixed specifications with temporal properties and state effects. Our entire framework has been implemented and formally verified in the Rocq Prover\footnote{Available at \artifacturl.}, yielding \emph{a reusable foundation for temporal reasoning over compositional effectful semantics}.

\paragraph{\bf Contributions.} Our main contributions are summarized as follows:
\begin{itemize}
    \item We propose \ibuchi{} -- as an instantiation of CTrees -- a compositional temporal specification language for characterizing comprehensive temporal behaviors of interaction trees.
    \item We develop trace- and tree-level satisfaction relations and compositional proof principles for verifying temporal properties of effectful computations.
    \item We integrate temporal specifications with the Dijkstra Monads Forever framework and provide a temporal encoding of generator-based future conditions.
  \item We implement and formally verify the proposed framework in the \rocq. Our implementation provides a basis for exploring temporal logic as a reusable specification component in the study of compositional program logics and semantics.
\end{itemize}

\paragraph{Paper Structure.}
\Cref{sec:moti-eg} motivates our approach with examples of specifying partially interpreted programs and of representing and
synthesizing future conditions; \cref{sec:bg} reviews the necessary background on
ITrees and CTrees. \Cref{sec:iauto} defines \ibuchi{} -- the domain of our temporal specifications -- together with its rich set of combinators, and \cref{sec:reasoning}
develops the reasoning system, comprising trace- and tree-level satisfaction rules as well as order theories for strengthening
and weakening \ibuchi{} specifications. \Cref{sec:dijkstra} embeds \ibuchi{} specifications into the Dijkstra Monads Forever framework and presents our
generator-based encoding of future conditions. \Cref{sec:cases} demonstrates the framework through several case studies. Finally, \cref{sec:related} discusses
related work and \cref{sec:conclusion} concludes with limitations and future directions.

\begin{figure}[t]
  \centering
  \begin{subfigure}{0.48\textwidth}
  \begin{minted}[fontsize=\footnotesize]{coq}
Definition guarded_alloc_store_free : com :=
  IF (var 1 <= 0) THEN SKIP
  ELSE   (* allocate memory into (ptr 0) *)
    ALLOC 0 IN
      IF (var 2 <= 0) THEN
        FREE (ptr 0)   (* free the memory *)
      ELSE
        (ptr 0 <<- 42);;      (* store 42 *)
        FREE (ptr 0).
  \end{minted}
  \caption{example program in \textsc{ImpMem}}
  \label{fig:example1}
  \end{subfigure}
  \hfil
  \begin{subfigure}{0.46\textwidth}
  \begin{tcolorbox}[enhanced, sharp corners, frame hidden, colback=black!0, fontupper=\itshape,
          borderline={0.5pt}{0pt}{black!60, decorate, decoration={random steps, segment length=4pt, amplitude=0.6pt}}]
			``Based on the return status of the memory allocation, the program in \Cref{fig:example1} should either (i) fail and report an error or (ii) succeed and finally free it after finitely many operations.''
    \end{tcolorbox}
  \caption{temporal spec.\ in natural language}
  \label{fig:example1-spec-nl}
  \end{subfigure}
  \caption{An example program in \textsc{ImpMem} (embedded in \rocq{}) and its temporal specification.}
\end{figure}

\section{Motivating Examples}
\label{sec:moti-eg}

Temporal properties describe how the observable events of a program unfold over time, e.g., whether one event eventually follows another or whether events always occur in a prescribed order.
Consider, for instance, the program in \Cref{fig:example1}, written in \textsc{ImpMem} -- a toy language extending \textsc{Imp} with sequential and mutable memory operations \cite{Pierce:SF1}. It invokes the system library functions \texttt{alloc} and \texttt{free} respectively for heap allocation and deallocation, whose implementations are intentionally left unspecified. Its intended (informal) specification is given in \Cref{fig:example1-spec-nl}.

Rather than designing a temporal program logic and verifying this property directly over the \textsc{ImpMem} source language, we first translate the program into an interaction tree of type 
\begin{align*}
\text{\coqtt{itree (stateE Env +' errorE +' fmemE nat)}}
\end{align*}%
where \coqtt{+'} denotes sum type for functors; \coqtt{stateE Env}, \coqtt{errorE}, and \coqtt{fmemE nat} are uninterpreted \emph{effects}.\footnote{\coqtt{Env} is the type of global variable-value mappings denoting the program state; \coqtt{nat} is the type of data stored in the heap.} Some of these effects correspond to implementation details rather than observable behaviors. For example, the \coqtt{stateE Env} effect can be interpreted via the standard state monad transformer \cite{liangMonadTransformersModular1995,tomschrijversMonadTransformersModular2019}. The \coqtt{errorE} effect could likewise be eliminated by the standard exception monad transformer, yet we retain it -- together with the \coqtt{fmemE nat} effect, which admits no interpretation -- as an observable behavior. After interpreting the state effect, the remaining ITree exposes only the observable interactions relevant to temporal reasoning, namely \coqtt{errorE +' fmemE nat}. Consequently, the original verification problem reduces to proving that this ITree -- parameterized by the current program state \coqtt{ρ: Env} -- satisfies the temporal property encoded in our \ibuchi{} type, as shown in \cref{fig:example1-spec}. See details of the \ibuchi{} type in \cref{sec:iauto}.

\begin{figure}[t]
  \centering
  \begin{minted}[fontsize=\footnotesize]{coq}
Definition guarded_alloc_store_free_spec (ρ : Env) : iBuchi (errorE +' fmemE nat) unit :=
  if val_env ρ 1 <=? 0 then nil else
    (alloc_success_event ++ if val_env ρ 2 <=? 0
        then free_attempt_event ++ nil
        else (store_attempt_event ⊎ read_attempt_event)✦ ++ free_attempt_event ++ nil)
    ⊎ (alloc_failure_event ++ report_error_event OutOfMemory ++ nil).

Theorem guarded_alloc_store_free_safe ρ
  : denote_state guarded_alloc_store_free ρ [tr⊨] guarded_alloc_store_free_spec ρ.
  \end{minted}
  \caption{Temporal specification of \Cref{fig:example1-spec-nl} in the \ibuchi{} type, together with the verification condition. Here, the \texttt{*\_event}s are shallow wrappers around the event constructors of the \ibuchi{} type; $\app$ denotes sequential concatenation of \ibuchi{} specifications; $\uni$ binary branching; $\finstar$ finite repetition; and $\&$ conjunction. The explicit \texttt{nil}s require the program to terminate at the corresponding points.}
  \label{fig:example1-spec}
\end{figure}

\begin{figure}[t]
  \centering
  \begin{minted}[fontsize=\footnotesize]{coq}
Variant all_infiniteF {E : Type → Type} {A : Type} (F : itree E A → Prop) : itree' E A → Prop :=
| MDivTau (t : itree E A) : F t → all_infiniteF F (TauF t)
| MDivVis (B : Type) (k : B → itree E A) (e : E B) : (∀ b, F (k b)) → all_infiniteF F (VisF e k).
Definition all_infinite_ {E A} (sim : itree E A → Prop) t := all_infiniteF sim (observe t).
Lemma all_infiniteF_mono {E A} (sim sim' : itree E A → Prop) t
      (IN : all_infiniteF sim t) (LE : sim <1= sim') : all_infiniteF sim' t.
Lemma all_infiniteF_mono' {E A} : monotone1 (@all_infinite_ E A).
Definition all_infinite {E A} := paco1 (@all_infinite_ E A) bot1.
  \end{minted}
  \caption{Example ITree predicate for defining divergence. To define such a coinductive predicate, one has to define a type-level function, to prove its monotonicity, and then to define it with the fixed-point operator, while performing pervasive lifting through the construction.}
  \label{fig:example1-itree-hard}
\end{figure}

\begin{remark}
Although existing definitions in the ITrees library \cite{itreeInfiniteCoqdoc} can express certain temporal-like properties of interaction trees, they do not provide a reusable, compositional framework for specifying and reasoning about complex temporal behaviors. For instance, \Cref{fig:example1-itree-hard} shows the boilerplate required to define divergence -- the property that every execution is infinite -- as a coinductive predicate using the Paco library \cite{hurPowerParameterizationCoinductive2013}. The issue becomes more pronounced for properties about the events emitted by an interaction tree: such hand-rolled specifications quickly grow cumbersome and are difficult to reuse or compose.
\qedT
\end{remark}

\begin{figure}[t]
  \centering
  \begin{minted}[fontsize=\footnotesize,escapeinside=@@]{coq}
Definition gen_memE_ftr {A} : EvAns (memE A) ~> iBuchi (memE A)
  := λ _ e, match e with
  | evempty _ _ _ => eme
  | evans _ MAlloc l =>
      @\textbf{F}@ (eve (λ e _, e = MFree l))
  | evans _ (MFree l) _ =>
      @\textbf{G}@ (eve (λ e _, e ≠ MFree l) & (eve (λ e _, e ≠ MGet l)) & (eve (λ e _, ∀ a, e ≠ MSet l a)))
  | evans _ (MGet l) _ => any
  | evans _ (MSet l _) _ => any
  end.
  \end{minted}
  \caption{Example of a future-condition generator that specifies memory access events.}
  \label{fig:example2}
\end{figure}


Our \ibuchi{} logic compositionally captures event-based temporal properties of ITrees; however, certain \emph{context-aware} properties are difficult to express directly. For the memory access events, an intended specification requires that \enquote{every allocated location is eventually freed, and no freed location is read, written, or freed again}. Such history-dependent specifications cannot be captured by our \ibuchi{} combinators alone. To address this limitation, we employ \emph{future-condition generators}, which synthesize temporal requirements on the future of a computation from the events observed so far. \Cref{fig:example2} exemplifies a future-condition generator for a simplified event type \coqtt{memE}, which differs from \coqtt{fmemE} in that its operations carry no failure status. This generator captures precisely the intended specification above. Unlike conventional temporal properties over program states, this specification naturally refers to interactions -- \coqtt{MFree l}, \coqtt{MAlloc} along with its result value, or other domain-specific events. Once the program is translated into an interaction tree, these behaviors become explicit observable events, so the specification can be expressed uniformly as a temporal property over the resulting tree, without introducing a dedicated logic for futures.

These examples illustrate a common verification workflow: programs from different source languages are first translated into interaction trees; language-specific internal effects are then interpreted away, leaving a semantic object that records only the observable interactions of interest. The remaining task is therefore independent of the source language: given an interaction tree, verify that its observable behaviors satisfy a temporal specification.
This observation motivates the development of \emph{a temporal proof system directly over interaction trees}. Once such a proof system is available, temporal reasoning becomes \emph{reusable} across all programming languages whose semantics is given in terms of interaction trees, while new programming abstractions can be supported simply by defining their observable events and corresponding effect
interpreters.

\section{Background}
\label{sec:bg}

\subsection{Interaction Trees}
\label{sec:itree}

%

Interaction Trees (ITrees) are a coinductive data structure for representing possibly nonterminating, effectful computations using algebraic effects and handlers~\cite{xiaInteractionTreesRepresenting2019}. They provide a reusable semantic framework for 
reasoning about interactive programs in the \rocq{}. 

\Cref{fig:itree-def} summarizes the core definitions of the ITrees library \cite{itreeInfiniteCoqdoc}. The coinductive type \coqtt{itree E R} is parameterized by an event signature \coqtt{E: Type -> Type} and a return type \coqtt{R}. An interaction tree is constructed from three primitives: \coqtt{Ret}, which terminates with a \emph{return} value; \coqtt{Tau}, which performs an \emph{internal} silent step; and \coqtt{Vis}, which issues an \emph{external} event together with a continuation describing the remaining computation. For every event signature \coqtt{E}, the type constructor \coqtt{itree E} forms a monad. Sequential computations are composed via the usual monadic operations \coqtt{ret} and \coqtt{bind}, while general recursive computations are expressed through the monadic iterator \coqtt{iter}. These combinators allow programs to be constructed \emph{compositionally}, in the same style as conventional monadic programming \cite{moggiComputationalLambdacalculusMonads1989,moggiNotionsComputationMonads1991,wadlerComprehendingMonads1992}.

Reasoning about interaction trees is performed modulo the \emph{weak bisimulation relation} \coqtt{eutt}, read ``equivalence up to \coqtt{Tau}'': two ITrees are \coqtt{eutt}-equivalent if they exhibit the same observable interactions and differ only in finitely many internal \coqtt{Tau} steps. This equational theory serves as the standard notion of semantic equivalence throughout the ITrees ecosystem.

The ITrees library is implemented on top of the Paco library~\cite{hurPowerParameterizationCoinductive2013}, which provides parameterized coinduction for defining and reasoning about coinductive structures. This proof infrastructure supports \emph{compositional coinductive proofs} without structuring proofs around \rocq{}'s syntactic guardedness condition for cofixpoints.

\begin{figure}[t]
  \centering
  \begin{minted}[fontsize=\footnotesize]{coq}
CoInductive itree (E : Type → Type) (R : Type): Type :=
| Ret (r : R)                                      (* computation terminating with value r *)
| Tau (t : itree E R)                              (* "silent" tau transition with child t *)
| Vis {A : Type} (e : E A) (k : A → itree E R). (* visible event e yielding an answer in A *)

Definition ret {E R} x : itree E R := Ret x.
Definition bind {E R S} (t : itree E R) (k : R → itree E S) : itree E S := (* omitted *)

Definition eutt {E A B} (R : A → B → Prop) : itree E A → itree E B → Prop := (* omitted *)

Lemma eutt_ret : RR r s → eutt RR (Ret r) (Ret s).
Lemma eutt_vis : (∀ a, eutt RR (k1 a) (k2 a)) → eutt RR (Vis e k1) (Vis e k2).
Lemma eutt_taulr : eutt RR t1 t2 → eutt RR (Tau t1) (Tau t2).
Lemma eutt_taul : eutt RR t1 t2 → eutt RR (Tau t1) t2.
Lemma eutt_taur : eutt RR t1 t2 → eutt RR t1 (Tau t2).

Notation "t1 ≈ t2" := (eutt eq).

Class MonadIter (M : Type → Type) `{Monad M} := iter : ∀ A B, (A → M (A + B)) → A → M B.

Instance iter_itree {E} : MonadIter (itree E) := (* omitted *)

Definition interp {E M : Type → Type} `{MonadIter M} {R : Type}
  (handler : ∀ A, E A → M A) : itree E R → M R := (* omitted *)
  \end{minted}
  \caption{Interaction trees: key types, operations and relations.}
  \label{fig:itree-def}
\end{figure}

\emph{Assigning Denotational Semantics to ImpMem.}
A key advantage of ITrees is that they can serve as a compositional denotational semantics for effectful programming languages. \Cref{fig:denotation} presents (an excerpt of) the denotational semantics of \textsc{ImpMem}, whose command syntax includes \coqtt{SKIP}, assignments, sequential composition, conditionals, and \coqtt{WHILE} loops. Rather than defining the execution of \textsc{ImpMem} programs as a transition relation, programs are denoted directly by an interaction tree where the state and memory operations are represented by visible interaction events with the environment: the \coqtt{CAss} command is denoted by \coqtt{State.Put} events (from \coqtt{stateE}), while the \coqtt{CAlloc}, \coqtt{CStore}, and \coqtt{CFree} commands are denoted by \coqtt{FMAlloc}, \coqtt{FMSet}, and \coqtt{FMFree} events (from \coqtt{fmemE}), respectively. The corresponding read operations are defined with the denotation of arithmetic expressions, \coqtt{denote_a} (omitted here), as the associated events \coqtt{State.Get} and \coqtt{FMGet} return values of type \coqtt{nat}. The remaining commands are denoted by the standard monadic operators; in particular, \coqtt{CWhile} is denoted via the \coqtt{iter} combinator introduced above. 

Once the syntax is given an ITree denotation, it is straightforward to further interpret the state events into an appropriate state monad. The type of \coqtt{handle_stateE} shows that the resulting computation lives in \coqtt{stateT Env (itree (errorE +' fmemE)) unit}, which unfolds to \coqtt{Env → itree (errorE +' fmemE) (Env * unit)}. The residual effects \coqtt{errorE +' fmemE} constitute the observable behaviors of the program; compositionally specifying and verifying their temporal properties is precisely the goal of this paper. 




\begin{figure}
  \centering
  \begin{minted}[fontsize=\footnotesize]{coq}
(* ImpMem commands *)
Inductive com :=
  | CSkip                                  (* SKIP         *)
  | CAss (x : Var) (a : aexp)              (* ::=          *)
  | CSeq (c1 c2 : com)                     (* ;;           *)
  | CIf (b : bexp) (ct ce : com)           (* IF THEN ELSE *)
  | CWhile (b : bexp) (body : com)         (* WHILE        *)
  | CAlloc (x : Var) (body : com)          (* ALLOC IN     *)
  | CStore (addr : pexp) (value : aexp)    (* <<-          *)
  | CFree (addr : pexp).                   (* FREE         *)

(* ImpMem Events *)
Definition ImpMemE : Type → Type := stateE Env +' errorE +' fmemE nat.

(* Denotation of ImpMem *)
Fixpoint denote_com (c : com) : itree ImpMemE unit :=
  match c with
  | CSkip => Ret tt
  | CAss x a => v <- denote_a a ;; ρ <- trigger (inl1 State.Get) ;; 
      trigger (inl1 (State.Put (update_val x v ρ)))
  | CSeq c1 c2 => denote_com c1 ;; denote_com c2
  | CIf b ct ce => guard <- denote_b b ;;
      if guard then denote_com ct else denote_com ce
  | CWhile b body => iter
      (fun _ => guard <- denote_b b ;;
        if guard then denote_com body ;; Ret (inl tt) else Ret (inr tt)) tt
  | CAlloc x body => ml <- trigger (inr1 (inr1 FMAlloc)) ;; match ml with
      | Some l => ρ <- trigger (inl1 State.Get) ;;
          trigger (inl1 (State.Put (update_loc x l))) ;; denote_com body
      | None => trigger (inr1 (inl1 (ReportError OutOfMemory)))
      end
  | CStore addr value => l <- denote_p addr ;; v <- denote_a value ;;
      trigger (inr1 (inr1 (FMSet l v))) ;; Ret tt
  | CFree addr => l <- denote_p addr ;;
      trigger (inr1 (inr1 (FMFree l))) ;; Ret tt
  end.

Definition handle_stateE
  : ∀ X, stateE X → stateT Env (itree (errorE +' fmemE nat)) X := (* omitted *)

Definition interp_ImpMem (t : itree ImpMemE unit)
  : stateT Env (itree (errorE +' fmemE nat)) unit := interp handle_stateE t.
  \end{minted}
  \caption{Denotational semantics of the \textsc{ImpMem} language using ITrees.}
  \label{fig:denotation}
\end{figure}

\subsection{Dijkstra Monads Forever}

While ITrees provide an expressive semantic representation of effectful programs, they do not by themselves offer a framework for verifying program properties. Dijkstra Monads Forever (DMF)~\cite{silverDijkstraMonadsForever2021} extends the Dijkstra monad methodology to ITrees, enabling weakest-precondition reasoning about possibly nonterminating interactive computations.

The central idea is to associate every computation monad with a corresponding \emph{specification monad} \cite{maillardDijkstraMonadsAll2019,swamyVerifyingHigherorderPrograms2013} and a \emph{monotone weakest-precondition transformer} \cite{dijkstraGuardedCommandsNondeterminacy1975} connecting the two. Specifications are thus themselves monadic objects, so that verification rules can be derived compositionally from the monadic structure of computations.

To support interactive computations, DMF develops specification monads over execution traces rather than over return values alone: Since an interactive computation may communicate with its environment indefinitely, specifications must capture both observable event sequences and termination behavior. Equipped with such trace-based specifications and with reasoning principles for iteration and interaction, the resulting framework supports \emph{weakest-precondition reasoning over ITrees} while remaining compatible with their compositional denotational semantics. 

However, DMF's specification language is fundamentally built upon pre- and post-conditions over traces. Although expressive for functional correctness and termination-sensitive properties, it does not directly support high-level temporal specifications, which motivates the development of dedicated temporal reasoning principles in this work. 


\subsection{Choice Trees}
\label{sec:ctree}

Choice Trees (CTrees) extend ITrees with native support for \emph{nondeterministic branching}, providing a compositional representation of effectful transition systems~\cite{chappeChoiceTreesRepresenting2023,chappeChoiceTreesRepresenting2025a}. CTrees enrich the three constructors of ITrees --- \coqtt{Ret}, \coqtt{Tau}, and \coqtt{Vis} --- with an additional branching constructor \coqtt{Br}\footnote{The library further provides \coqtt{Guard} and \coqtt{Stuck}, corresponding to unary and nullary \coqtt{Br}, respectively, as well as \coqtt{Step}, which performs an observable $\tau$-transition. Note that the CTree counterpart of ITrees' \coqtt{Tau} is \coqtt{Guard}, not \coqtt{Step}.}. Unlike \coqtt{Vis}, whose successor is determined by an observable response from the environment, \coqtt{Br} represents an explicit branching point over a family of continuations, so that both internal and external nondeterministic choices are modeled directly as part of the semantic object.

CTrees retain the monadic interface of ITrees and thus support the same compositional construction of effectful computations. The additional branching constructor, however, enables nondeterminism to be represented natively rather than encoded through effects or auxiliary semantic layers, making branching behavior a first-class component of the underlying semantic model.

Instead of relying primarily on an equational theory of \coqtt{eutt}-style weak bisimulation, the CTrees library adopts a \emph{transition-based view of computation}: The operational behavior of a choice tree is described by a labeled transition system induced by its constructors, over which a hierarchy of behavioral relations --- strong similarities or strong/weak bisimilarities --- is defined coinductively. These transition-based relations support both refinement and equivalence reasoning in the presence of nondeterministic branching, while remaining compatible with the compositional structure inherited from ITrees. 

The CTrees library is implemented using the Coinduction library~\cite{damienpousCoinductionAllWay2016}, which provides a generic infrastructure for defining greatest fixed points and conducting coinductive proofs over complete lattices. Together with the native support for nondeterministic branching, this proof infrastructure renders CTrees a convenient basis for constructing, in the remainder of this paper, richer semantic objects for temporal specifications with branching behavior.
\section{Automata-Based Temporal Logic}
\label{sec:iauto}

This section presents our \ibuchi{} type -- the domain of temporal specifications for ITrees -- whose definition is inspired by the way the \itree{} machinery is reused to define the \itrace{} type \cite{silverDijkstraMonadsForever2021,itraceDefinitionCoqdoc}. Intuitively, \itree{}s correspond to computation trees and \itrace{}s to the linear streams refining them; the \ibuchi{} type then corresponds to \emph{sets of such streams}, playing a role similar to that of classical \emph{regular expressions}, yet capable of expressing both finite and infinite properties. Hence, instead of building a \emph{deep embedding} of an assertion language for temporal specifications, we represent temporal formulas more directly via a \emph{shallow embedding} into a dedicated domain type -- the \ibuchi{} type\footnote{Although the term \emph{B\"uchi automaton} customarily refers to automata over infinite words, whose acceptance requires visiting designated states infinitely often, our \ibuchi{} type encodes both terminating and diverging properties of program traces.} in our terminology. 

Before defining the \ibuchi{} type, we first recap the \itrace{} type~\cite{silverDijkstraMonadsForever2021}, both because it provides useful intuition for our definition and because the satisfaction relation between \itree{}s and \ibuchi{} temporal formulas is built on top of \itrace{}s.

\begin{figure}
  \centering
  \includegraphics[width=0.8\linewidth]{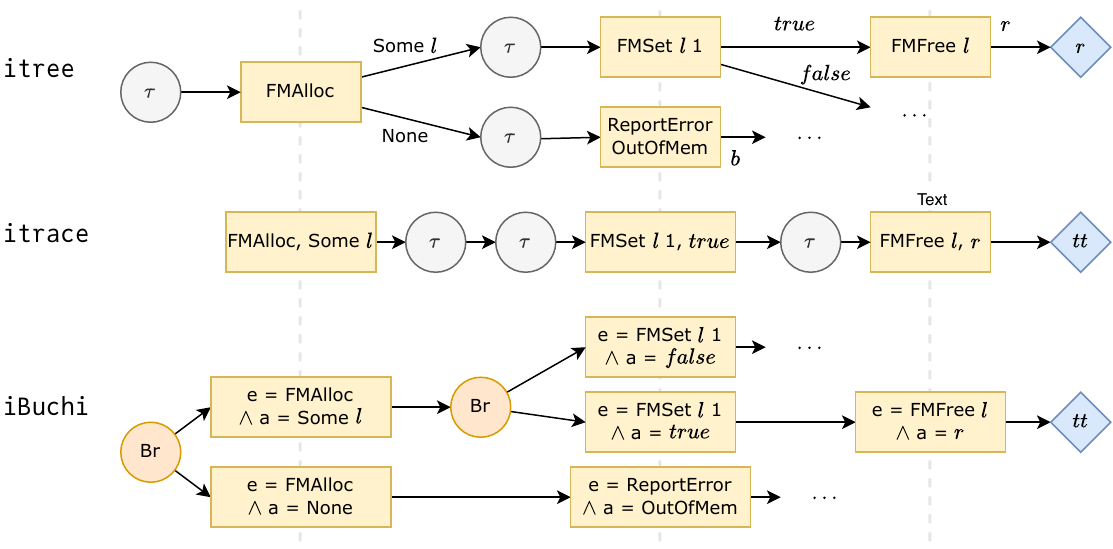}
  \caption{%
    Structural comparison of \itree{}s, \itrace{}s, and \ibuchi{} specifications, with observable events aligned across the three
    representations. The depicted \itrace{} refines the \itree{}: each observable event of the \itree{} corresponds, in the \itrace{},
    to a pair of the event itself and an answer value. Moreover, not only the depicted one but \emph{every} \itrace{} refining the
    \itree{} satisfies the \ibuchi{} specification at the bottom.
  }
  \label{fig:itree-itrace-ibuchi}
\end{figure}

\subsection{ITrace: An Intermediate Abstraction Layer}

To provide a precise and algebraically well-behaved specification monad for \itree{}s with uninterpreted events, \citet{silverDijkstraMonadsForever2021} proposed an intermediate abstraction layer --- the \itrace{} type --- defined directly in terms of the \itree{} type itself. Like \itree{}s, the \itrace{} type is parameterized by an event type \coqtt{E} and a return type \coqtt{R}; it is defined as an interaction tree over the following auxiliary \coqtt{EvAns} event type:
\begin{minted}[fontsize=\footnotesize,xleftmargin=2em]{coq}
Variant EvAns (E : Type → Type) : Type → Type :=
  | evans : ∀(A : Type) (ev : E A) (ans : A), EvAns E unit
  | evempty : ∀(A : Type) (ev : E A) (Hempty : A → void), EvAns E void.

Definition itrace (E : Type → Type) (R : Type) := itree (EvAns E) R.
\end{minted}
Intuitively, an event of type \coqtt{EvAns E} packs an event \coqtt{ev: E A} with either (i) a possible answer \coqtt{ans: A}, where its \coqtt{unit} return type admits a continuation with a unique branch, or (ii) the evidence that \coqtt{A} is empty, where its \coqtt{void} return type admits the continuation over the empty domain. These restrictions make such special \itree{}s deterministic: an \itrace{} is essentially a linear, branch-free computation path. With the \coqtt{trace_refine} relation ($\sqsubseteq$) defined below, the set of all \itrace{}s refining a given \itree{} captures \emph{all} information about the tree.
\begin{minted}[fontsize=\footnotesize,xleftmargin=2em]{coq}
Variant event_refine {E : Type → Type} : ∀ (A B : Type), EvAns E A → E B → Prop :=
  | rer {A : Type} (e : E A) (a : A) : event_refine A (evans e a) e
  | ree {A : Type} (e : E A) (Hempty : A → void) : event_refine A (evempty e Hempty) e.

Variant event_refine_cont {E : Type → Type}
    : ∀ (A B : Type), EvAns E A → A → E B → B → Prop :=
  | rar {A : Type} (e : E A) (a : A) : event_refine_cont (evans e a) tt e a.

Definition trace_refine {E R} := @euttEv (EvAns E) E R R event_refine event_refine_cont eq.
Notation "b ⊑ t" := (trace_refine b t) (at level 90).
\end{minted}
\Cref{fig:itree-itrace-ibuchi} illustrates this intuition with an \itree{} instance and an \itrace{} refining it; the arbitrarily placed \iTau{} nodes indicate that rewriting modulo \coqtt{eutt} is permitted on both sides of $\sqsubseteq$.


\subsection{\ibuchi{}: Tree-Shaped Automata}
\label{sec:ibuchi-def}

We now develop the domain of our temporal logic, the \ibuchi{} type. To this end, we define a new event type, \coqtt{EvRx}, analogous to \coqtt{EvAns} except that it matches \emph{sets of\, \coqtt{EvAns} events} rather than single ones. We first define the \coqtt{Char} type:
\begin{minted}[fontsize=\footnotesize,xleftmargin=2em]{coq}
Inductive Char (E : Type → Type) : Type :=
  | ChrPrd : (∀ R, E R → R → Prop) → Char E
  | ChrWld : Char E
  | ChrNon : Char E
  | ChrAnd : Char E → Char E → Char E
  | ChrOr  : Char E → Char E → Char E.
\end{minted}
The \coqtt{Char} type serves as a small domain-specific language for matching individual answered events -- in analogy with character classes in regular expressions: its constructors represent predicates over \coqtt{EvAns} events, the wildcard, the empty set, intersection, and union, respectively. Augmenting \coqtt{Char} with an additional constructor that matches \coqtt{EvAns} events carrying the \coqtt{void} answer type, we define the \coqtt{EvRx} event type as follows:
\begin{minted}[fontsize=\footnotesize,xleftmargin=2em]{coq}
Variant EvRx (E : Type → Type) : Type → Type :=
  | RxEme : EvRx E void
  | RxChr : Char E → EvRx E unit.
\end{minted}

Finally, we define the \ibuchi{} type as \ctree{}s with the \coqtt{EvRx} event type. As for the choice indices in the definition below, \coqtt{Index.B2} is the type of binary choices, used for branching specifications, while \coqtt{Index.BN} is the type of countably many choices, which is essential for encoding the temporal property of \emph{finite repetition} (see the Star combinators in \cref{sec:ibuchi-constructs}):
\begin{minted}[fontsize=\footnotesize,xleftmargin=2em]{coq}
Definition iBuchi E R := ctree (EvRx E) (Index.B2 +' Index.BN) R.
\end{minted}

\subsection{\ibuchi{} Constructs}
\label{sec:ibuchi-constructs}

Next, we equip the \ibuchi{} type with the standard combinators. \Cref{fig:ibuchi-constructs} presents their \rocq{} definitions; we describe the major ones below.

\begin{figure}[t]
  \centering
  \begin{minted}[fontsize=\footnotesize,escapeinside=@@]{coq}
Definition nil {E} : iBuchi E unit := Ret tt.

Variant lift_prd {E T} (p : E T → T → Prop) : ∀ R, E R → R → Prop :=
  mk_lift_prd : ∀ e r, p e r → lift_prd p e r.

Definition eve {E R} (p : E R → R → Prop) : iBuchi E unit
  := CTree.trigger (RxChr (ChrPrd (lift_prd p))).

Definition wc {E} : iBuchi E unit := CTree.trigger RxWld.

Definition uni {E R} (t f : iBuchi E R) : iBuchi E R := br2 t f.
Infix "⊎" := uni (at level 50, left associativity).

Definition app {E R S} (r : iBuchi E R) (s : iBuchi E S) : iBuchi E S
  := r >>= (λ _, Guard (s)).
Infix "++" := app (at level 60, right associativity).

CoFixpoint ω {E} (r : iBuchi E unit) : iBuchi E unit := nil ⊎ (r ++ ω r).
Notation "r  '⋆'" := (ω r) (at level 75).

CoFixpoint ω_inf {E} (r : iBuchi E unit) : iBuchi E unit
  := CTree.bind r (λ _, Guard (ω_inf r)).
Notation "r  '⊛'" := (ω_inf r) (at level 75).

Definition any {E} : iBuchi E unit := wc⋆.

CoFixpoint neg {E R} (r : iBuchi E R) : iBuchi E R := (* omitted *)
Notation "'¬'  r" := (neg r) (at level 75).

Fixpoint rep {E} (n : nat) (rx : iBuchi E unit) : iBuchi E unit
  := match n with
  | O => nil
  | S n' => rx ++ rep n' rx
  end.

Definition star {E} (rx : iBuchi E unit) : iBuchi E unit := brN (λ n, rep n rx).
Notation "r  '✦'" := (star r) (at level 75).

CoFixpoint prd {E} (r s : iBuchi E unit) : iBuchi E unit := (* omitted *)
Infix "&" := prd (at level 69, left associativity).

(* LTL Modalities *)
Definition @\textbf{G}@ {E} (r : iBuchi E unit) : iBuchi E unit := r⋆.
Definition @\textbf{U}@ {E} (r s : iBuchi E unit) : iBuchi E unit := r✦ ++ s.
Definition @\textbf{W}@ {E} (r s : iBuchi E unit) : iBuchi E unit := r⋆ ++ s.
Definition @\textbf{F}@ {E} (r : iBuchi E unit) : iBuchi E unit := @\textbf{U}@ (¬ r) r ++ any.
Definition @\textbf{X}@ {E} (r : iBuchi E unit) : iBuchi E unit := wc ++ r ++ any.
  \end{minted}
  \caption{The \ibuchi{} combinators.}
  \label{fig:ibuchi-constructs}
\end{figure}


\paragraph{Termination} The \nil{} combinator is a thin wrapper around the \coqtt{Ret} constructor of the \ctree{} type. When used alone, i.e., with no formula following it, \nil{} marks the termination point of the preceding \ibuchi{} formula.

\paragraph{Sequencing} The $\app$ combinator wraps the \coqtt{bind} ($\bind$) operator of the \ctree{} monad, discarding the value passed to the continuation. It thus serves as the sequential composition of \ibuchi{} specifications.

\paragraph{Observable Events} The \eve{} combinator turns a predicate over events and their answers into an atomic \ibuchi{} formula via the \coqtt{ChrPrd} constructor. Since a user-supplied predicate typically concerns a specific return type \coqtt{T: Type}, i.e., has type \coqtt{E T → T → Prop}, the auxiliary inductive type \coqtt{lift_prd} lifts it to the dependent form \coqtt{∀ R, E R → R → Prop} required by \coqtt{ChrPrd}. We also provide the \neg{} operator for \coqtt{Char}-wise negation; see its detailed definition in \cref{sec:omitted-defs}.

\paragraph{Branching} Since \ctree{}s are designed precisely to internalize nondeterministic choice, their unobservable branching constructor \coqtt{Br} (\cref{sec:ctree}) directly encodes the branching of temporal formulas. The $\uni$ combinator instantiates \coqtt{Br} with the binary choice index, allowing a formula to proceed with either of the two operands.

\paragraph{Stars} The \ibuchi{} type admits \emph{three} distinct repetition combinators:
\begin{itemize}
\item $r\ \finstar$, defined via the \coqtt{rep} combinator, repeats the formula $r$ a \emph{finite} number of times, akin to the
classical semantics of the \emph{Kleene star}.
\item $r\ \mostar$ repeats $r$ \emph{infinitely} many times, with the option to \emph{terminate} after each round;
\item $r\ \infstar$ repeats $r$ \emph{infinitely} many times, without any option to terminate;
\end{itemize}

\paragraph{Products} The intersection operation on events makes it possible to define a novel construct -- the \emph{product} (or, viewed logically, the \emph{conjunction}) of two \ibuchi{} objects. Products allow two or (finitely) more \ibuchi{} formulas to be imposed simultaneously, which is a key ingredient in our encoding of future conditions. The detailed definition of the $\prd$ combinator is deferred to \cref{sec:omitted-defs}.

\paragraph{LTL Modalities} With the above combinators, the standard modalities of \emph{Linear Temporal Logic} (LTL) can be derived naturally. The \emph{Globally} modality $\mathbf G\ r$ is given by iteration as $r\ \mostar$, requiring $r$ throughout, whether or not the program terminates. The \emph{Until} (resp.\ \emph{Weak Until}) modality sequences a finite (resp.\ possibly infinite) iteration prefix with a target formula. The \emph{Finally} modality is then derived from Until: $\mathbf F\ r$ permits a finite prefix violating $r$ before $r$ is reached. Similarly, the \emph{Next} modality $\mathbf X\ r$ consumes a single arbitrary event before continuing with $r$.

\begin{remark} 
    One may wonder how temporal properties of observed events, as encoded by \ibuchi{}, compare in expressiveness with traditional state-based propositional LTL formulas. A direct way to bridge the two settings is to enrich the interpretation of the operations of interest so that they emit dedicated \coqtt{Probe} events, each carrying the current program state. Over the resulting \itree{}, state-based temporal properties can then be specified as \ibuchi{} formulas referring to these \coqtt{Probe} events; in this sense, our event-based setting subsumes state-based reasoning.
\qedT 
\end{remark} 

\section{Reasoning System}
\label{sec:reasoning}

This section presents the satisfaction relations of our \ibuchi{}-based temporal logic.
We first define when an \itrace{} satisfies an \ibuchi{} formula and present the low-level
reasoning rules for this trace-level satisfaction.
We then lift this notion to a satisfaction relation between \itree{}s and \ibuchi{} formulas
and derive the corresponding tree-level reasoning rules.
Finally, we develop preorder theories for the \ibuchi{} type, enabling strengthening and
weakening of \ibuchi{} formulas.

\subsection{Trace-Level Satisfaction}
\label{sec:reasoning:tc-sat}

The trace satisfaction relation $\sat$ is the lowest-level semantic relation for \ibuchi{} formulas, upon which the subsequent relations are built. We start with individual events: The \coqtt{SatRx} relation defines when an \itrace{} event (of type \coqtt{EvAns}) satisfies an \ibuchi{} event (of type \coqtt{EvRx}), where the auxiliary relation \coqtt{CharSat} matches events carrying nonempty answers against \coqtt{Char} patterns:
\begin{minted}[fontsize=\footnotesize,xleftmargin=2em]{coq}
Inductive CharSat E : EvAns E unit → Char E → Prop :=
  (** match an event satisfying the predicate *)
  | sc_prd {A} (e : E A) (a : A) (per : ∀ R, E R → R → Prop)
    : per A e a → CharSat (evans A e a) (ChrPrd per)

  (** match any possible event *)
  | sc_wld (e : EvAns E unit) : CharSat e ChrWld

  (** no rule for [ChrNon] *)

  (** match conjunctions *)
  | sc_and (e : EvAns E unit) (r1 r2 : Char E)
    : CharSat e r1 → CharSat e r2 → CharSat e (ChrAnd r1 r2)

  (** match disjunctions *)
  | sc_orl (e : EvAns E unit) (r1 r2 : Char E)
    : CharSat e r1 → CharSat e (ChrOr r1 r2)
  | sc_orr (e : EvAns E unit) (r1 r2 : Char E)
    : CharSat e r2 → CharSat e (ChrOr r1 r2).

Inductive SatRx (E : Type → Type) : ∀ (A B : Type), EvAns E A → EvRx E B → Prop :=
  | rxe_e {A} (e : E A) (Hempty : A → void)
    : SatRx (evempty A Hempty e) RxEme         (* match an empty event *)
  | rxe_c (e : EvAns E unit) (c : Char E)
    : CharSat e c → SatRx e (RxChr c).         (* match non-empty events *)
\end{minted}

The \coqtt{SatRx} relation induces a matching relation \coqtt{SatL} on transition labels: besides \coqtt{obs} labels carrying \coqtt{SatRx}-related events, it also relates any pair of \coqtt{val} labels, thereby requiring the trace and the temporal formula to terminate simultaneously: 
\begin{minted}[fontsize=\footnotesize,xleftmargin=2em]{coq}
Inductive SatL E : @label (EvAns E) → @label (EvRx E) → Prop :=
  | satL_ev {R} {ea : EvAns E R} {ra} {eb : EvRx E R} {rb}
    : SatRx ea eb → SatL (obs ea ra) (obs eb rb)
  | satL_ret {A B} {r : A} {s : B}
    : SatL (val r) (val s).
\end{minted}

Finally, the $\sat$ relation is defined as the \ctree{} \emph{strong similarity} relation (\coqtt{ssim}) instantiated with the label relation \coqtt{SatL} above:
\begin{minted}[fontsize=\footnotesize,xleftmargin=2em]{coq}
Definition sat E A B := @ssim (EvAns E) (EvRx E) Index.B0 (Index.B2 +' Index.BN) A B (@SatL E).
Notation "x  '⊨'  y" := (@sat _ _ _ x y) (at level 50, no associativity).
\end{minted}

Expanding the underlying coinductive definition of \ctree{} strong similarity, $\sat$ can be understood as the largest heterogeneous relation $SAT$ such that,
whenever a \coqtt{ctrace}\footnote{Although the $\sat$ relation is ultimately meant to relate the \itrace{} and the \ibuchi{} type, it is more convenient to establish this strong similarity over the \coqtt{ctrace} type (\ctree{}s with \coqtt{EvAns} events and nullary choice).
We keep the \embed{} function explicit to indicate that some satisfaction rules apply only to canonical \itrace{}s. As mentioned in \cref{sec:ctree}, \coqtt{Tau} constructors of ITrees are mapped to \coqtt{Guard} constructors, so \coqtt{eutt}-equivalent ITrees are mapped to strongly bisimilar \ctree{}s.
}~\coqtt{E A} emits a visible event in \coqtt{EvAns E} and transitions to its tail, the \ibuchi{} formula also emits an event in \coqtt{EvRx E} and transitions to its tail, with the two events related by \coqtt{SatL} and the two tails related again by $SAT$.

\begin{figure}[t]
  \begin{mathpar}
    \inferrule[Sat-Nil]{ }{\Ret\ a \sat \nil}
    \and
    \inferrule[Sat-App]{t \sat r \\ u \sat s}{(t\, ;\!; u) \sat r \app s}
    \and
    \inferrule[Sat-Eutt]{t \approx u \\ t \sat r}{u \sat r}
    \\
    \inferrule[Sat-Any]{ }{\embed\ t \sat \any}
    \and
    \inferrule[Sat-Weaken]{t \sat r \\ r \ibsimchr s}{t \sat s}
    \and
    \inferrule[Sat-Eve]
        {P\ e\ a}
        {\trigger\ (\text{\coqtt{evans}}\ e \ a) \sat \eve\ P}
    \\
    \inferrule[Sat-Eve-Neg]
        {\neg P\ e\ a}
        {\trigger\ (\text{\coqtt{evans}}\ e \ a) \sat \neg (\eve\ P)}
    \and
    \inferrule[Sat-Eve-NegT]
        {a : A \\ P : E\ B\to B\to \text{\coqtt{Prop}} \\ A \ne B}
        {\trigger\ (\text{\coqtt{evans}}\ e \ a) \sat \neg (\eve\ P)}
    \\
    \inferrule[Sat-Prd]{t \sat r \\ t \sat s}{t \sat r \prd s}
    \and
    \inferrule[Sat-Prd-Elim-L]{t \sat r \prd s}{t \sat r}
    \and
    \inferrule[Sat-Prd-Elim-R]{t \sat r \prd s}{t \sat s}
    \\
    \inferrule[Sat-Uni-L]{t \sat r}{t \sat r \uni s}
    \and
    \inferrule[Sat-Uni-R]{t \sat s}{t \sat r \uni s}
    \and
    \inferrule[Sat-Uni-Elim]{t \sat r \uni s \\ t = \Ret\ a\ \lor\ t = \Vis\ e\ k}{t \sat r\quad \lor\quad t \sat s}
    \\
    \inferrule[Sat-$\star$-Nil]{t \sat \nil}{t \sat r\mostar}
    \and
    \inferrule[Sat-$\star$-Step]{t \sat r \app r\mostar}{t \sat r\mostar}
    \and
    \inferrule[Sat-$\infstar$-Step]{t \sat r \app r\infstar}{t \sat r\infstar}
    \\
    \inferrule[Sat-$\finstar$-Nil]{t \sat \nil}{t \sat r\finstar}
    \and
    \inferrule[Sat-$\finstar$-Rep]{n : \mathbb{N} \\ t \sat \rep\ n\ r}{t \sat r\finstar}
    \and
    \inferrule[Sat-$\finstar$-Elim]
        {\Vis\ e\ k \sat r\finstar}
        {∃\ n : \mathbb{N} .\ t \sat \rep\ n\ r}
  \end{mathpar}
  \caption{The trace-level satisfaction rules ($t \sat rx$).}
  \label{fig:tc-rules}
\end{figure}

The above strong similarity relation captures both finite and infinite properties of traces via different \ibuchi{} specifications on its right-hand side; however, certifying such a relation always by directly expanding the underlying coinductive definition is intricate. We therefore develop a set of reasoning rules for proving trace satisfaction, in line with the \ibuchi{} combinators of \Cref{fig:ibuchi-constructs}. 

\Cref{fig:tc-rules} collects these rules, all of which have been mechanized in \rocq. In the rule \textsc{Sat-Weaken}, the preorder $\ibsimchr$ is the \emph{Character-Aware Refinement} (see \cref{sec:reasoning:order}). In practice, stronger preorders or even equivalences are also used to transform the right-hand side of the $\sat$ relation. It is also worth noting that the elimination rules for \coqtt{Br}-based \ibuchi{} combinators (i.e., \textsc{Sat-Uni-Elim} and \textsc{Sat-$\finstar$-Elim}) cannot be proved in their general forms, since it is not possible to choose a branch without taking an explicit transition. In practice, however, the restricted versions for \coqtt{Vis} and \coqtt{Ret} suffice, combined with \textsc{Sat-Eutt} to erase the leading \coqtt{Tau}s.

\subsection{Tree-Level Satisfaction}
\label{sec:reasoning:tr-sat}
The trace satisfaction relation naturally induces the ITree satisfaction relation:

\begin{minted}[fontsize=\footnotesize,xleftmargin=2em]{coq}
Definition trSat {E R T} (tr : itree E R) (rx : iBuchi E T) : Prop
  := ∀ tc, tc ⊑ tr → embed tc ⊨ rx.
Notation "tr  '[tr⊨]'  rx" := (trSat tr rx)  (at level 50, no associativity).
\end{minted}

\begin{figure}[t]
  \begin{mathpar}
    \inferrule[TrSat-Nil]{ }{\Ret~ r \trSat \nil}
    \and
    \inferrule[TrSat-Eutt]{t \approx u \\ t \trSat r}{u \trSat r}
    \and
    \inferrule[TrSat-Any]{ }{t \trSat \any}
    \and
    \inferrule[TrSat-Weaken]{ t \trSat r \\ r \ibsimchr s}{t \trSat s}
    \\
    \inferrule[TrSat-Bind]{t \trSat r \\ \forall a.\ k\ a \trSat s}{t \bind k \trSat r \app s}
    \and
    \inferrule[TrSat-Bind-Trigger]
        {P\ e\ a \\ \forall a.\ P\ e\ a \to k\ a \trSat r}
        {\trigger\ e \bind k \trSat \eve\ P \app r}
    \and
    \inferrule[TrSat-Bind-Trigger-Neg]
        {A\ \text{is inhabited} \\ \forall a\! :\! A.\ \lnot P\ e\ a \\ \forall a.\ k\ a \trSat r}
        {\trigger\ e \bind k \trSat (\lnot\eve\ P) \app r}
    \and
    \inferrule[TrSat-Bind-Trigger-NegT]
        {A\ \text{is inhabited} \\ P : E\ B \to B \to \text{\coqtt{Prop}} \\ A \ne B \\ \forall a.\ k\ a \trSat r}
        {\trigger\ e \bind k \trSat (\lnot\eve\ P) \app r}
    \\
    \inferrule[TrSat-Prd]{t \trSat r \\ t \trSat s}{t \trSat r \prd s}
    \and
    \inferrule[TrSat-Uni-L]{t \trSat r}{t \trSat r \uni s}
    \and
    \inferrule[TrSat-Uni-R]{t \trSat s}{t \trSat r \uni s}
    \\
    \inferrule[TrSat-$\star$-Nil]{t \trSat \nil}{t \trSat r\mostar}
    \and
    \inferrule[TrSat-$\star$-Step]{t \trSat r \app r\mostar}{t \trSat r\mostar}
    \and
    \inferrule[TrSat-$\finstar$-Nil]{t \trSat \nil}{t \trSat r\finstar}
    \and
    \inferrule[TrSat-$\finstar$-Rep]{n : \mathbb{N}\\ t \trSat \rep\ n\ r}{t \trSat r\finstar}
    \\
    \inferrule[TrSat-$\infstar$-Step]{t \trSat r \app r \infstar}{t \trSat r\infstar}
    \and
    \inferrule[TrSat-Iter-Nonempty]
        {\forall i.\ \body\ i \trSat r \\ \forall i\ b.\ b \sqsubseteq \body\ i \to b \not\approx \Ret\ \_}
        {\iter\ \body\ i \trSat r\mostar}
    \\
    \inferrule[TrSat-Iter-Nonempty-Noexit]
        {\forall i.\ \body\ i \trSat r
            \\ \forall i\ b.\ b \sqsubseteq \body\ i \to b \not\approx \Ret\ \_ \land \lnot \text{\coqtt{may_converge}}\ (\text{\coqtt{inr}}\ \_)\ b}
        {\iter\ \body\ i \trSat r\infstar}
  \end{mathpar}
  \caption{The tree-level satisfaction rules ($t \trSat r$).}
  \label{fig:tr-rules}
\end{figure}

Building on the lower-level trace satisfaction rules, we develop the ITree satisfaction rules listed in \Cref{fig:tr-rules} -- all proved in the \rocq.
Most of them are direct counterparts of the trace rules.
Here, we sketch the proofs of two non-trivial rules, corresponding to the two key combinators of ITrees, \coqtt{bind} and \coqtt{iter}, respectively.\footnote{All the lemmas used in the proofs are provided in \cref{sec:appx-lemma-in-proofs}.} We begin with \textsc{TrSat-Bind}, which exposes a structural compatibility between \ibuchi{} temporal formulas and \itree{} sequential composition via \coqtt{bind}.

\begin{theorem}[\textsc{TrSat-Bind}]
\begin{mathpar}
  \inferrule
    {t \trSat r \\ \forall a.\ k\ a \trSat s}
    {t \bind k \trSat r \app s}~.
\end{mathpar}
\end{theorem}

\begin{proof}[Proof sketch]
Fix any refinement $tc ⊑ t \bind k$. By \coqtt{decompose_trace_refine_bind} from the ITree trace library, we decompose $tc$ as $tc_1 \bind g$ with $tc_1 ⊑ t$. Thus $tc_1$ captures the prefix produced by $t$, while $g$ is the residual continuation. From the assumption $t \trSat r$, we obtain $\embed\ tc_1 \sat r$. The remainder of the proof proceeds by distinguishing whether $tc_1$ converges.

If $tc_1$ converges to some \(a\), then \coqtt{trace_refine_bind_cont_inv} from the ITree trace library shows that the residual continuation at the return point still refines the original one, i.e., $g\ a ⊑ k\ a$. Hence $\embed\ (g\ a)$ satisfies \(s\) by the assumption on \(k\). We then split the overall trace into the prefix $tc_1$ and the suffix $g\ a$: the lemma \coqtt{sat_Ret_prefix} removes the trivial \coqtt{Ret} continuation from the prefix, and the \textsc{Sat-App} rule combines the satisfaction of the prefix and suffix to conclude $\embed\ tc \sat r \app s$.

If $tc_1$ is infinite, then the continuation is never semantically reached. It therefore suffices to show that $\embed\ (tc_1 \bind g)$ satisfies $r \app s$ using only the fact that $\embed\ tc_1 \sat r$. This is proved by a coinductive argument, packaged in \coqtt{sat_bind_divergent}: \coqtt{trans_embed_val_inf_false} shows that an embedded infinite trace cannot perform a value transition, ruling out any actual entry into the continuation, while \coqtt{trans_embed_obs_inf_inf} shows that every observable step leads again to an embedded infinite suffix, so the coinduction hypothesis can be reapplied. Therefore, $\embed\ tc \sat r \app s$ holds in the divergent case as well.
\end{proof}

We next illustrate the proof pattern behind our $\iter$ rules with \textsc{TrSat-Iter-Nonempty}. The proof of \textsc{TrSat-Iter-Nonempty-Nonexit} is structurally similar and hence omitted here.

\begin{theorem}[\textsc{TrSat-Iter-Nonempty}]
\begin{mathpar}
  \inferrule
    {\forall i.\ \body\ i \trSat r \\ \forall i\ tc\ a.\ tc \sqsubseteq \body\ i \to tc \not\approx \Ret\ a}
    {\iter\ \body\ i \trSat r\mostar}~.
\end{mathpar}
\end{theorem}

\begin{proof}[Proof sketch]
Consider an arbitrary refinement $tc ⊑ \iter\ \body\ i$. After unfolding one step of $\iter$, we decompose $tc$ as $tc_1 \bind g$ with $tc_1 ⊑ \body\ i$. Thus $tc_1$ is the current loop-body segment and $g$ is the residual loop continuation. By the assumption that the loop body satisfies $r$ for every input, we obtain $\embed\ tc_1 \sat r$. The proof then unfolds $r\mostar$ and distinguishes whether $tc_1$ converges.

If $tc_1$ converges to some result, the non-emptiness assumption rules out the case where $tc_1$ is equivalent to an immediate $\Ret$; hence the current iteration contributes a genuine $r$-segment. By \coqtt{trace_refine_bind_cont_inv} from the ITree trace library, the residual continuation reached after this return still refines the corresponding continuation of the original iterator. If the result is \coqtt{inl}~$al$, the loop continues, and the coinduction hypothesis shows that the remaining suffix satisfies $r\mostar$. If the result is \coqtt{inr}~$ar$, the loop exits, and the suffix is discharged by the empty branch of the star. In both subcases, \coqtt{sat_Ret_prefix} recovers satisfaction of the current segment, while \coqtt{SatL_is_update_val_rel} connects the return-insensitive satisfaction of the segment with the return-sensitive reasoning about the continuation.

If $tc_1$ is infinite, the argument is analogous to the divergent case of \textsc{TrSat-Bind}: the continuation is never semantically reached, so it suffices to show that binding the infinite prefix preserves the unfolded $r\mostar$ specification. Since $\embed\ tc_1 \sat r$, this follows again from \coqtt{sat_bind_divergent}.
\end{proof}

\subsection{\ibuchi{} Order Theories}
\label{sec:reasoning:order}









Next, we present several \emph{order relations} on the \ibuchi{} type. The semantic order is given by trace inclusion, while the proof-oriented orders are based on strong (bi)similarity and its character-aware variant.  These relations play different roles: trace inclusion states the intended semantic meaning of weakening, whereas the (bi)similarity-based orders provide compositional proof principles.

\subsubsection{Semantic Trace Inclusion}
\label{sec:order:inc}

The most direct semantic ordering is given by trace inclusion.  We write \(P \ibtinc Q\) when every concrete trace satisfying \(P\) also satisfies \(Q\):
\[
P \ibtinc Q
\ddefine
\forall t\colon\ \ t \sat P \,\Rightarrow\, t \sat Q~.
\]
Accordingly, \(P\) is stronger than \(Q\), while \(Q\) is a semantic weakening of \(P\). This relation captures specification refinement extensionally, solely in terms of the sets of traces admitted by the two specifications. As usual, it induces the corresponding semantic equivalence:
\[
P \ibtinceq Q
\ddefine
P \ibtinc Q \,\land\, Q \ibtinc P~.
\]

\subsubsection{Structural Refinement}



Beyond trace inclusion, the CTree structure of \ibuchi{} specifications induces a proof-oriented order, called \emph{structural refinement}, based on strong similarity. We write
\[
P \ibsim Q
\]
when \(P\) strongly simulates \(Q\), using Leibniz equality as the relation on labels.  Operationally, every transition of \(P\) must be matched by a corresponding transition of \(Q\), after which the resulting continuations must again be related by simulation. This relation supports \emph{structural and compositional proofs}: $P \ibsim Q$ means that $P$ is a stronger specification and can therefore be weakened to $Q$.
Note that structural refinement is a preorder and thus induces the equivalence relation
\[
P \ibsimeq Q
\ddefine
P \ibsim Q \,\land\, Q \ibsim P~.
\]

\subsubsection{Character-Aware Refinement}


Although structural refinement, based on Leibniz equality of labels, often suffices, it can be overly restrictive in how observable events are matched. A natural relaxation is to first define an order on \coqtt{Char}, i.e., \coqtt{CharLe}, and then lift it to a weaker notion of strong similarity on \ibuchi{} specifications. This yields the \emph{character-aware refinement}, written
\[
P \ibsimchr Q~.
\]
\begin{minted}[fontsize=\footnotesize,xleftmargin=2em]{coq}
Definition CharLe {E : Type → Type} : relation (Char E)
  := λ c d, ∀ e, CharSat e c → CharSat e d.

Inductive RxLabelLe (E : Type → Type)
  : @label (EvRx E) → @label (EvRx E) → Prop :=
  | rxl_tau : RxLabelLe τ τ
  | rxl_val {A} (a : A) : RxLabelLe (val a) (val a)
  | rxl_chr (c d : Char E) : CharLe c d → RxLabelLe (obs (RxChr c) tt) (obs (RxChr d) tt).

Definition iBuchiLeChar {E R} : relation (iBuchi E R)
  := λ r s, @ssim (EvRx E) (EvRx E) _ _ R R (@RxLabelLe E) r s.
Infix "[≲χ]" := iBuchiLeChar (at level 90, no associativity).
\end{minted}
In this variant of strong similarity, labels are compared using \coqtt{RxLabelLe} rather than Leibniz equality. In particular, observable character labels are related according to semantic inclusion. Consequently, a specification containing a more precise \coqtt{Char}-pattern may be weakened to one containing a more general pattern.

Character-aware refinement is again a preorder and therefore induces the equivalence relation
%
%
\[
P \ibsimchreq Q
\ddefine
P \ibsimchr Q \,\land\, Q \ibsimchr P~.
\]

\subsubsection{Bisimilarity and Character-Aware Bisimilarity}

Finally, we consider strong bisimilarity based on Leibniz equality and its character-aware variant, written
\[
P \ibbisim Q \qand P \ibbisimchr Q~,
\]
respectively. Unlike $\ibsim$ and $\ibsimchr$, they are not only preorders but also equivalence relations defined directly by \emph{strong bisimilarity}. Strong bisimilarities are similar to strong similarities, except that the transition-matching condition is symmetric. The $\ibbisim$-based lemmas come directly from the CTrees library, allowing simpler equational reasoning about \ibuchi{} formulas.

\subsubsection{Relationships Between the Orders}

We now establish the relationships between the proposed orders.
First, the proof-oriented orders are sound with respect to semantic trace inclusion:
\[
P \ibsim Q
\Longrightarrow
P \ibtinc Q\,,
\qand
P \ibsimchr Q
\Longrightarrow
P \ibtinc Q~.
\]
The first implication states that ordinary structural refinement preserves trace satisfaction, while the second extends the result to character-aware weakening.

Second, the Leibniz equality-based relations imply their character-aware counterparts, namely
\begin{align*}
  P \ibsim Q
\,\Longrightarrow\,
P \ibsimchr Q~,
\qquad
P \ibsimeq Q
\,\Longrightarrow\,
P \ibsimchreq Q~,
\qquad
P \ibbisim Q
\,\Longrightarrow\,
P \ibbisimchr Q~.
\end{align*}


Third, each strong bisimilarity implies the corresponding equivalence induced by refinement:
\[
P \ibbisim Q
\Longrightarrow
P \ibsimeq Q\,,
\quad\text{and}
\quad P \ibbisimchr Q
\Longrightarrow
P \ibsimchreq Q~.
\]

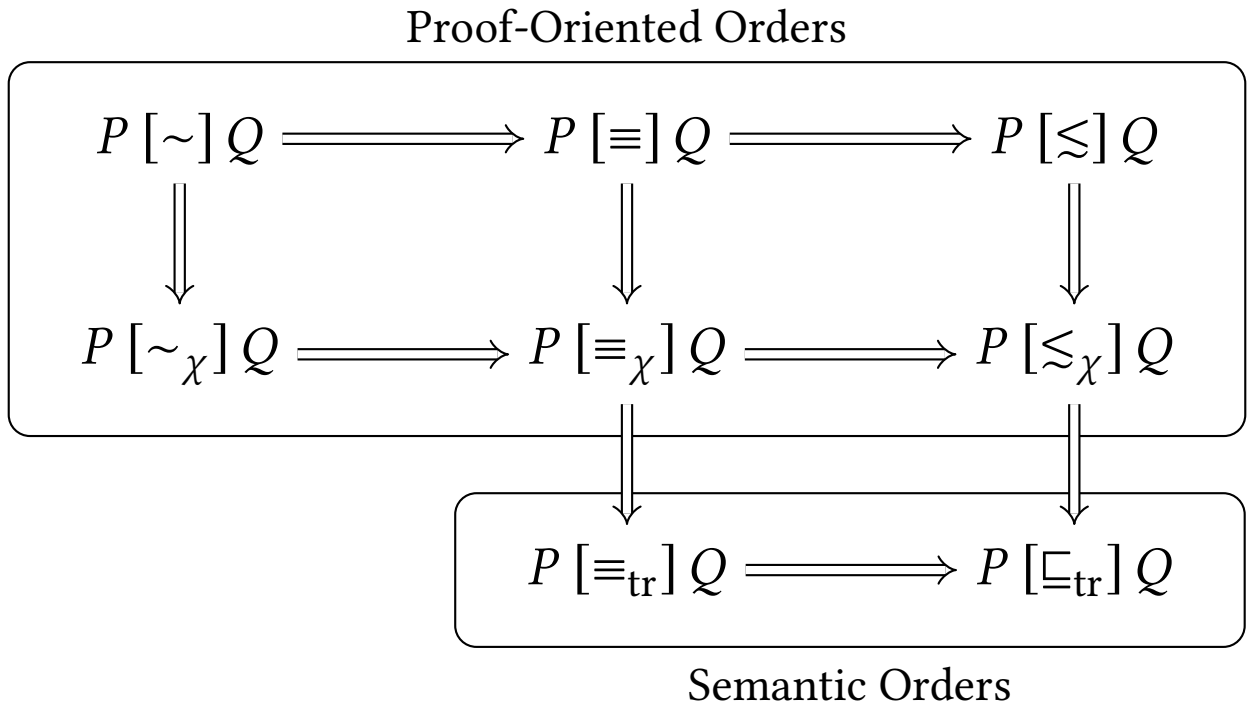
\begin{figure}[t]
\centering
\begin{tikzpicture}[
  >=Latex,
  node distance=0.8cm and 1.6cm,
  impl/.style={double equal sign distance, -Implies},
  bgbox/.style={
    draw,
    rounded corners,
    inner xsep=1em,
    inner ysep=0.6em
  }
]
\node (ibbisim) {$P \ibbisim Q$};
\node[right=of ibbisim] (ibsimeq) {$P \ibsimeq Q$};
\node[right=of ibsimeq] (ibsim) {$P \ibsim Q$};
\node[below=of ibbisim] (ibbisimchr) {$P \ibbisimchr Q$};
\node[below=of ibsim] (ibsimchr) {$P \ibsimchr Q$};
\node[below=of ibsimeq] (ibsimchreq) {$P \ibsimchreq Q$};
\node[below=of ibsimchr] (ibtinc) {$P \ibtinc Q$};
\node[below=of ibsimchreq] (ibtinceq) {$P \ibtinceq Q$};

\draw[impl] (ibbisim) -- (ibsimeq);
\draw[impl] (ibbisim) -- (ibbisimchr);
\draw[impl] (ibsimeq) -- (ibsim);
\draw[impl] (ibsim) -- (ibsimchr);
\draw[impl] (ibsimeq) -- (ibsimchreq);
\draw[impl] (ibbisimchr) -- (ibsimchreq);
\draw[impl] (ibsimchreq) -- (ibsimchr);
\draw[impl] (ibsimchr) -- (ibtinc);
\draw[impl] (ibsimchreq) -- (ibtinceq);
\draw[impl] (ibtinceq) -- (ibtinc);

\begin{scope}[on background layer]
  \node[
    bgbox,
    fit=(ibbisim) (ibsimeq) (ibsim)
        (ibbisimchr) (ibsimchreq) (ibsimchr),
    label={[font=\small]above:Proof-Oriented Orders}
  ] {};
  \node[
    bgbox,
    fit=(ibtinceq) (ibtinc),
    label={[font=\small]below:Semantic Orders}
  ] {};
\end{scope}

\end{tikzpicture}
\caption{%
Implication relationships among semantic and proof-oriented orders on \ibuchi{} specifications. 
}
\label{fig:ibuchi-orders}
\end{figure}

We do not rely on any of the converse implications above. In particular, deriving similarity or bisimilarity from semantic trace inclusion would amount to establishing a completeness result for the proof-oriented relations, which we do not assume here. \Cref{fig:ibuchi-orders} summarizes all the implications established above and provides an overview of the relationships among these order theories. All the proofs of these relationships have been mechanized in \rocq.

\section{Future Conditions with \ibuchi{} Logic}
\label{sec:dijkstra}



In this section, we show how our \ibuchi{}-based temporal specifications can be used to instantiate the recently proposed framework of future conditions \cite{songProveNFixTemporalPropertyGuided2024,songSpecifyingVerifyingFuture2026}.
As a main contribution of DMF, the \coqtt{TraceSpec} specification type~\cite{silverDijkstraMonadsForever2021} provides a generic interface for reasoning about finite execution prefixes and their possible continuations. We instantiate this interface with \ibuchi{} specifications, using \ibuchi{} formulas as temporal predicates over traces while keeping the underlying \coqtt{TraceSpec} interface independent of any trace logic system.

\subsection{Encoding \ibuchi{} Specifications in TraceSpec}
\label{sec:encode-ibuchi-tracespec}

A temporal specification at a given execution state naturally consists of two components. The first constrains the \emph{history observed so far}, while the second constrains the \emph{possible future continuations} of that history. We thus encode a pair of \ibuchi{} specifications as a \coqtt{TraceSpec} assertion\footnote{Here, \coqtt{↑} embeds a finite list of \coqtt{EvAns} events into an \coqtt{itrace} and \coqtt{+++} denotes concatenation of \coqtt{itrace}s.}:
\begin{minted}[fontsize=\footnotesize,xleftmargin=2em]{coq}
Definition encode {A} (pre : iBuchi E unit) (future : iBuchi E unit) : TraceSpec E A
  := λ log p, embed (↑ log) ⊨ pre
    ∧ ∀ b, embed (↑ log +++ b) ⊨ future
      → p (↑ log +++ b).
\end{minted}
Intuitively, the history specification \coqtt{pre} is required to hold on the execution prefix produced so far; The future specification \coqtt{future}, by contrast, restricts the continuations relevant to the continuation predicate: For every possible continuation $b$, if the extended trace satisfies \coqtt{future}, then the continuation predicate must hold on that extended trace.

This separation reflects the operational structure of \coqtt{TraceSpec}: the current execution prefix is already fixed, whereas its future continuation remains abstract. The future specification therefore acts as an \emph{assumption} restricting the admissible continuations. Moreover, the encoding does not require \coqtt{TraceSpec} to inspect the internal structure of \ibuchi{} formulas; it relies only on their satisfaction relation over traces.

Importantly, the future specification is interpreted over the \emph{entire extended trace} rather than over the continuation suffix alone. Thus, if $log$ is the current prefix and $b$ is a continuation, the future formula is evaluated on $log\ \text{\coqtt{+++}}\ b$. This allows future specifications to refer uniformly to both past observations and future events, without requiring a separate relative satisfaction judgment.

\subsection{Generating Future Conditions from Events}
\label{sec:generating-future-conditions}

Although users may provide future specifications explicitly, doing so is often too low-level. Many temporal obligations are naturally associated with individual events. For example, a request event may require a later response, a lock event may require a later unlock, and an output event may require some future acknowledgment. Rather than requiring users to compose all such obligations manually, we \emph{generate} future conditions from the events encountered during execution.

We assume an event generator \coqtt{gen: EvAns E ~> iBuchi E}%
\footnote{Here, \coqtt{E ~> F} for \coqtt{E F: Type → Type} expands to \coqtt{∀ X, E X → F X}.}
that maps each observed trace event to an \ibuchi{} future-condition formula specifying the required behavior \emph{after} that event occurs. Intuitively, generation over a trace would be defined by the following \coqtt{CoFixpoint}:

\begin{minted}[fontsize=\footnotesize,xleftmargin=2em]{coq}
Fail CoFixpoint gen_spec (gen : EvAns E ~> iBuchi E) (ct : ctrace E unit) : iBuchi E unit
  := match observe ct with
  | RetF _ => Ret tt
  | StuckF => Stuck
  | StepF t => Step (gen_spec gen t)
  | GuardF t => Guard (gen_spec gen t)
  | BrF c k => Stuck
  | VisF (evempty _ _ _) _ => eme
  | VisF ((evans _ e a) as ev) k =>
      (eve (λ e' a', e' = e ∧ a' = a) ++ (gen_spec gen (k tt)) & gen _ ev)
  end.
\end{minted}
However, \rocq{} rejects this definition because it violates the guardedness condition for cofixpoint definitions.\footnote{The recursive occurrence of \coqtt{gen_spec} appears beneath the non-constructor $\prd$. \rocq{} conservatively requires every recursive occurrence in cofixpoints to be guarded by a constructor \cite{rocq-prover-coinductive-9-2}.}
%
To overcome this limitation, given such a generator, we define a mapping from a finite list of events, represented by the \coqtt{ev_list E} type, to the accumulated future specification:

\begin{minted}[fontsize=\footnotesize,xleftmargin=2em]{coq}
Fixpoint gen_spec (gen : EvAns E ~> iBuchi E) (el : ev_list E) : iBuchi E unit
  := match el with
  | [] => any
  | (evempty _ _ _) :: _ => eme
  | ((evans _ e a) as ev) :: el' =>
      eve (λ e' a', e' = e ∧ a' = a) ++ (gen_spec gen el' & gen _ ev)
  end.
\end{minted}
The empty trace generates the trivial specification \coqtt{any}, which is satisfied by every trace, since no event has introduced a future obligation. For a non-empty trace, the generated specification first records the occurrence of the head event and then imposes two requirements: the obligations generated from the remaining events and the local obligation introduced by the head event itself. These two specifications are combined using the $\prd$ operation.

Thus, each event contributes a local temporal requirement, while the generated specification accumulates all such requirements along the trace. Future-condition generation is therefore event-local at the interface level, yet yields a global temporal specification for the entire execution.

Finally, using the finite-prefix generator \coqtt{gen_spec}, we instantiate \coqtt{TraceSpec} without requiring an explicit future formula. The user supplies a history specification and an event generator; the future condition is then derived from \emph{every} finite event prefix of a continuation:
\begin{minted}[fontsize=\footnotesize,xleftmargin=2em]{coq}
Definition encode_ftr {A} (pre : iBuchi E unit) (gen : EvAns E ~> iBuchi E) : TraceSpec E A
  := λ log p, embed (↑ log) ⊨ pre
    ∧ ∀ b : itrace E A,
      (∀ el : ev_list E, ∀ t,
        (↑ log +++ b) ≈ ↑ el ;; t
        → embed (↑ log +++ b) ⊨ gen_spec gen el)
      → p (↑ log +++ b).
\end{minted}

Operationally, this encoding requires the current history to satisfy the history specification. For every admissible continuation, the extended trace must also satisfy the future condition generated from each of its finite event prefixes. In the special case where the history specification is \coqtt{nil}, future-condition generation is applied from an empty history. For a more direct presentation of this case, we isolate the generated future-condition obligation as the following property:
\begin{minted}[fontsize=\footnotesize,xleftmargin=2em]{coq}
Definition generated_ftr_vc {A} (gen : EvAns E ~> iBuchi E) (program : itree E A) : Prop
  := ∀ tc : itrace E A, tc ⊑ program
    → ∀ el : ev_list E, ∀ t,
      tc ≈ ↑ el ;; t
      → embed tc ⊨ gen_spec gen el.
\end{minted}

Despite the complexity of the event-triggered future-condition generation, the resulting concise \coqtt{generated_ftr_vc} property for an empty history admits the following compositional reasoning lemmas: one for terminating programs and one for a head event followed by its continuation.
\begin{minted}[fontsize=\footnotesize,xleftmargin=2em]{coq}
Lemma generated_ftr_vc_ret {A}
    (gen : EvAns E ~> iBuchi E) (a : A)
  : generated_ftr_vc gen (Ret a).

Lemma generated_ftr_vc_bind_trigger {A B}
    (event : E A) (k : A → itree E B) (gen : EvAns E ~> iBuchi E)
  : (∃ a : A, True)
  → (∀ a, generated_ftr_vc gen (k a))
  → (∀ a, k a [tr⊨] gen _ (evans _ event a))
  → generated_ftr_vc gen
      (ITree.bind (ITree.trigger event) k).
\end{minted}

\section{Implementation and Case Studies}
\label{sec:cases}

We have fully mechanized the proposed framework in Rocq (\artifacturl). The formalization covers the \ibuchi{} type and its combinators, trace- and tree-level satisfaction relations and their associated reasoning rules, order-theoretic results for \ibuchi{} specifications, and generator-based future conditions. The core Rocq development contains $\sim$5,000 lines of definitions and proofs, excluding examples and case studies. Below, we demonstrate through several case studies that our proof system for temporal specifications and future conditions is not only sufficiently expressive for real-life programs but also convenient to work with.

\subsection{Terminating and Diverging}
\label{sec:cases:term-div}
We show that our proof system captures the \emph{termination-sensitive} nature of the \itree{} semantics. 
Specifically, we illustrate how \ibuchi{} specifications can be used to express the \emph{finally} and \emph{globally} LTL modalities and then combine them into a single specification:
\begin{minted}[fontsize=\footnotesize,xleftmargin=2em,escapeinside=@@]{coq}
Definition inc_trace : ctrace (ioE nat) unit
  := (cofix inc n := CTree.bind (CTree.trigger (evans _ (Output n) tt))
        (λ _, Guard (inc (n + 1)))) 0.

Definition finally_ge3_spec : iBuchi (ioE nat) unit
  := @\textbf{F}@ (eve (λ e _, ∃ x, e = Output x ∧ x >= 3)).

Definition wc_then_globally_ge1_spec : iBuchi (ioE nat) unit 
  := wc ++ @\textbf{G}@ (eve (λ e _, ∃ x, e = Output x ∧ x >= 1)).

Example inc_prd_sats : inc_trace ⊨ finally_ge3_spec & wc_then_globally_ge1_spec.
\end{minted}
The \coqtt{finally_ge3_spec} specification precisely captures traces that eventually trigger an \coqtt{Output} event carrying an argument no less than $3$, while no preceding \coqtt{Output} event carries such an argument. Moreover, when the argument of $\finstar$ in the definition of \textbf{F} denotes only finite traces -- as in this case, where it is the negation of a single-event specification -- the final matching event is required to occur.
The \coqtt{wc_then_globally_ge1_spec} specification states that, after an initial event ignored by the wildcard, every subsequent \coqtt{Output} event must carry an argument no less than $1$.
After proving these two specifications separately, we can apply the \textsc{Sat-Prd} rule to combine the proofs and derive satisfaction of the composite specification.


It is worth noting that, when a program violates a termination-sensitive specification, the corresponding satisfaction claim can be refuted. For example, a \Ret{} \coqtt{ctrace} cannot satisfy an infinite repetition $\infstar$ of wildcards, whereas \coqtt{inc_trace} cannot satisfy any finite repetition of wildcards:
\begin{minted}[fontsize=\footnotesize,xleftmargin=2em,escapeinside=@@]{coq}
Example ret_not_ω_inf_wc
  : (Ret tt : ctrace (ioE nat) unit) ⊨ (wc ⊛) → False.

Example inc_trace_not_finite_rep_wc (n : nat)
  : inc_trace ⊨ rep n wc → False.
\end{minted}

These examples demonstrate that our specifications distinguish termination from divergence precisely, rather than merely approximating observable event behavior.

\subsection{The Malloc-Free Game}
\label{sec:cases:mf-game}


\Cref{sec:moti-eg} has illustrated concrete scenarios in which compositional temporal specifications are essential. In this section, we model \texttt{malloc}, \texttt{free}, and memory-access operations as uninterpreted events, and verify an example future specification in detail within our framework. We first define \coqtt{memE}, a simplified version of the \coqtt{fmemE} event family as introduced in \Cref{sec:moti-eg}:
\begin{minted}[fontsize=\footnotesize,xleftmargin=2em,escapeinside=@@]{coq}
Definition Loc := nat.      (* use natural numbers to represent memory locations *)
Variant memE (A : Type) : Type → Type :=
  | MAlloc                  : memE A Loc
  | MFree (l : Loc)         : memE A unit
  | MGet  (l : Loc)         : memE A A
  | MSet  (l : Loc) (a : A) : memE A unit.

Definition gen_memE_ftr {A} : EvAns (memE A) ~> iBuchi (memE A)
  := λ _ e, match e with
  | evempty _ _ _ => eme
  | evans _ MAlloc l => @\textbf{F}@ (eve (λ e _, e = MFree l))
  | evans _ (MFree l) _ =>
    @\textbf{G}@ (eve(λ e _, e ≠ MFree l) & (eve(λ e _, e ≠ MGet l)) & (eve(λ e _, ∀ a, e ≠ MSet l a)))
  | evans _ (MGet l) _ => any
  | evans _ (MSet l _) _ => any
  end.
\end{minted}

Intuitively, \coqtt{memE (A: Type): Type → Type} is an event family parameterized by a type \coqtt{A}, representing the type of values stored in memory. In this case study, we instantiate \coqtt{A} with \coqtt{nat} for simplicity. The event constructors of \coqtt{memE A} have the following meanings:
\begin{itemize}
\item \coqtt{MAlloc} represents memory allocation and returns a location of type \coqtt{Loc}. For simplicity, we omit allocation failure and similarly ignore failure cases for the remaining operations; hence, the return type is not wrapped in \coqtt{option}.
\item \coqtt{MFree (l: Loc)} represents deallocation of location \coqtt{l} and returns \coqtt{unit}.
\item \coqtt{MGet (l: Loc)} represents reading from location \coqtt{l} and returns a value of type \coqtt{A}.
\item \coqtt{MSet (l: Loc) (a: A)} represents writing the value \coqtt{a} to location \coqtt{l} and returns \coqtt{unit}.
\end{itemize}


\begin{figure}[t]
\centering
\begin{minted}[fontsize=\footnotesize,xleftmargin=2em,escapeinside=@@]{coq}
Definition maybe_managed_loop : itree (memE nat) unit :=
  (* Outer polling loop. The environment controls [MGet 0]. Answer 0 keeps the loop
  running, so this prefix may diverge before allocation. Any non-zero answer exits
  the loop and starts the managed region. *)
  ITree.iter (fun _ : unit =>
      (n <- ITree.trigger (MGet 0) ;;
       iRet (if n =? 0 then inl tt else inr tt)))
    tt ;;
  initial <- ITree.trigger (MGet 1) ;;
  (* Once the polling loop exits, allocation definitely happens. Henceafter, the program
  must eventually free the allocated location to satisfy the generated future obligation. *)
  l <- ITree.trigger MAlloc ;;
  (* Inner bounded countdown loop. For every natural number [initial], the counter is
     decremented until [0]; then the program frees [l] and exits. *)
  ITree.iter (fun remaining : nat =>
      (ITree.trigger (MSet l remaining) ;;
       iRet (if remaining =? 0
          then inr tt
          else inl (Nat.pred remaining))))
    initial ;;
  ITree.trigger (MFree l).

Theorem maybe_managed_loop_ftr_safe
  : verify_cond _ (encode_true_ftr nil gen_memE_ftr) maybe_managed_loop.
\end{minted}
\caption{Example \itree{} program with memory operation events and its verification condition.}
\label{fig:memE-itree-eg}
\end{figure}

For the future-condition generator, we reuse the one from \Cref{sec:moti-eg}. We aim to verify the future conditions for the program of type \coqtt{itree (memE nat) unit} shown in \Cref{fig:memE-itree-eg}. The proof follows the syntactic structure of the program, decomposing it into three layers: an outer polling prefix, a managed region beginning with \coqtt{MAlloc}, and a post-allocation countdown loop ending with \coqtt{MFree l}.
Before proving the generated verification condition (VC) for the whole program, we use the following lemma to handle the two loops. Since the single-event body of each loop generates only a trivial future condition, the lemma reduces the VC of the loop to that of its continuation:
\begin{minted}[fontsize=\footnotesize,xleftmargin=2em,escapeinside=@@]{coq}
Lemma generated_ftr_vc_iter_trigger {I R A B} (event : I → E A) (next : I → A → I + R)
    (k : R → itree E B) (gen : EvAns E ~> iBuchi E) (i : I)
  : (∃ a : A, True) → (∀ i a, gen _ (evans _ (event i) a) = any)
  → (∀ r, generated_ftr_vc gen (k r))
  → generated_ftr_vc gen (ITree.bind (ITree.iter
      (λ i, ITree.bind (ITree.trigger (event i)) (λ a, iRet (next i a))) i) k).
\end{minted}

The proof proceeds from the outermost program fragment to the innermost one, following the successive reductions of the generated VC by the proof rules and auxiliary lemmas. The initial goal is the VC for the whole program. The first reduction applies \coqtt{generated_ftr_vc_iter_trigger} to the following prefix:
\begin{minted}[fontsize=\footnotesize,xleftmargin=2em,escapeinside=@@]{coq}
ITree.iter (fun _ : unit =>
  (n <- ITree.trigger (MGet 0) ;;
  iRet (if n =? 0 then inl tt else inr tt))) tt ;;
initial <- ITree.trigger (MGet 1) ;;
...
\end{minted}

This step reduces the proof to the continuation reached when the loop exits. The polling loop may diverge if the environment continues to return \coqtt{0}, but such divergence occurs before \coqtt{MAlloc} and therefore does not violate any future obligation introduced by allocation. The remaining goal begins with reading the initial counter:
\begin{minted}[fontsize=\footnotesize,xleftmargin=2em,escapeinside=@@]{coq}
...
initial <- ITree.trigger (MGet 1) ;;
l <- ITree.trigger MAlloc ;;
...
\end{minted}

This visible event is handled by the generic lemma \coqtt{generated_ftr_vc_bind_trigger}, reducing the goal to the continuation for an arbitrary value of \coqtt{initial}. After this reduction, the remaining program fragment begins with allocation. The next step applies the allocation rule to \coqtt{MAlloc}:
\begin{minted}[fontsize=\footnotesize,xleftmargin=2em,escapeinside=@@]{coq}
...
l <- ITree.trigger MAlloc ;;
ITree.iter (fun remaining : nat =>
  ...
\end{minted}

Applying \coqtt{generated_ftr_vc_bind_trigger} again introduces a fresh location \coqtt{l} and reduces the goal to the VC of the post-allocation continuation. At the same time, it generates the essential temporal side condition that the continuation must satisfy
\[
\_ \quad\trSat \quad \text{\coqett{@\textbf{F}@ (eve (λ e _, e = MFree l))}}~.
\]

The remaining VC therefore concerns the following countdown loop and its continuation:
\begin{minted}[fontsize=\footnotesize,xleftmargin=2em,escapeinside=@@]{coq}
...
ITree.iter (fun remaining : nat =>
  (ITree.trigger (MSet l remaining) ;;
  iRet (if remaining =? 0 then inr tt else inl (Nat.pred remaining)))) initial ;;
ITree.trigger (MFree l)
\end{minted}
The goal reduces to two facts about this fragment. First, the fragment itself satisfies the generated VC by the lemma \coqtt{generated_ftr_vc_iter_trigger}. Second, it satisfies the eventual-free property \coqett{[tr⊨] @\textbf{F}@ (eve (λ e _, e = MFree l))}, because, for any value of \coqtt{initial}, the counter eventually decreases to \coqtt{0}, after which the exit branch immediately performs \coqtt{MFree l}. The final \coqtt{MFree l} event additionally generates the temporal side condition that no further operation is performed on \coqtt{l}; this condition is trivially satisfied by the empty continuation.


\subsection{Mixed Specification with State Effect and Temporal Logic}
\label{sec:cases:state}

The original proof system for future conditions proposed by \citet{songProveNFixTemporalPropertyGuided2024, songSpecifyingVerifyingFuture2026} combines traditional Hoare-style pre- and post-conditions with temporally defined future conditions, yielding an expressive reasoning system for program verification. Below, we show that Hoare-style conditions, formulated as predicates over program states, and our generator-based future conditions can be represented as orthogonal, modular components within a unified framework -- the Dijkstra Monad Forever framework -- while inheriting the algebraic structure and extensibility of DMF.

\begin{figure}[t]
\centering
\begin{minted}[fontsize=\footnotesize,xleftmargin=2em,escapeinside=@@]{coq}
Definition account_prog : itree (stateE nat +' memE nat) nat :=
  n <- ITree.trigger (inl1 (State.Get nat)) ;;
  l <- ITree.trigger (inr1 MAlloc) ;;
  ITree.trigger (inl1 (State.Put nat (2 * n + 1))) ;;
  m <- ITree.trigger (inl1 (State.Get nat)) ;;
  ITree.trigger (inr1 (MFree l)) ;;
  ITree.trigger (inl1 (State.Put nat (S m))) ;;
  iRet n.

(* Program State Pre-Condition *)
Definition account_pre (s : nat) : Prop :=
  2 <= s ∧ Nat.even s = true.

(* Program State Post-Condition *)
Definition account_post : StateRelPost nat nat :=
  fun s s' ret => s' = S (2 * s + 1) ∧ ret = s ∧ Nat.even s' = true.

(* Program Future Condition Generator *)
Definition account_future_gen : FutureGenerator nat (memE nat) :=
  fun _ => gen_memE_ftr.

(* StateFutureSpec Embedding *)
Definition account_spec : StateFutureSpec nat (memE nat) nat :=
  StateFutureSpec.encode_mixed account_pre account_future_gen account_post.
\end{minted}
\caption{Encoding state-based pre- and post-conditions together with generator-based future conditions.}
\label{fig:mixed-eg}
\end{figure}

In \Cref{fig:mixed-eg}, we show how a mixed specification consisting of pre-conditions, post-conditions, and generator-based future conditions can be encoded in the \coqtt{StateFutureSpec} type. This type is defined as \coqtt{StateSpecT S (TraceSpec E)}, obtained by applying the \coqtt{StateSpecT} monad transformer with state type \coqtt{S} to the \coqtt{TraceSpec E} monad over the event type \coqtt{E} of the target semantics. Intuitively, an element of this type has the underlying type \coqtt{S → TraceSpecInput E (S * A)}, together with the structure and laws required for it to form a specification monad.
The following unfolding lemma makes explicit how these conditions are encoded in \coqtt{StateFutureSpec}:
\begin{minted}[fontsize=\footnotesize,xleftmargin=2em,escapeinside=@@]{coq}
Lemma verify_encode_mixed_unfold {A}
    (spre : S → Prop) (gen : FutureGenerator) (post : StateRelPost A) (m : StateFuture A)
  : verify_cond (encode_mixed spre gen post) prog
   ↔ ∀ s b, spre s → b ⊑ prog s →
     (∀ s' a, may_converge (s', a) b → post s s' a)
     ∧ (∀ el t, b ≈ append (ev_list_to_stream el) t
        → embed b ⊨ FutureSpec.gen_spec (gen s) el).
\end{minted}
Assuming the pre-condition holds, every trace \coqtt{b} refining the execution of \coqtt{prog} from the initial state must satisfy the following two requirements: (1) if \coqtt{b} terminates, then its final state and return value satisfy the post-condition; and (2) \coqtt{b} satisfies the generator-based future verification condition.

Consider the program in \Cref{fig:mixed-eg} of type \coqtt{itree (stateE nat +’ memE nat) nat}. It first reads the initial account state \coqtt{s}, allocates a memory location, updates the state to \coqtt{2 * s + 1}, reads this intermediate state, frees the allocated location, increments the state once more, and finally returns the initial value \coqtt{s}.
The pre-condition requires the initial account state to be at least \coqtt{2} and even. The post-condition is defined over the initial state \coqtt{s}, the final state \coqtt{s’}, and the return value \coqtt{ret}. It requires that \coqtt{s' = S (2 * s + 1)}, \coqtt{ret = s}, and the final state \coqtt{s'} is even. Although the future-condition generator may in general depend on the initial state, this example uses the constant generator \coqtt{gen_memE_ftr}; see \Cref{sec:cases:mf-game}.
Finally, \coqtt{encode_mixed} combines the pre-/post-conditions and future-condition generator into the mixed specification \coqtt{account_spec}. The resulting VC can then be discharged using the reasoning rules and auxiliary lemmas introduced above.

\section{Related Work}
\label{sec:related}

\paragraph{Interaction Trees}
\citet{xiaInteractionTreesRepresenting2019} propose \emph{Interaction Trees} -- a coinductive variant of the free monad \cite{swierstraDataTypesCarte2008} -- as an expressive semantic domain for effectful and possibly diverging computations in Gallina, the pure and terminating specification language of the Rocq Prover \cite{rocq-prover-refman-9-2}. ITrees come with rich equational theories, e.g., \emph{equivalence up to \texttt{Tau}} (\coqtt{eutt}), which abstracts away finitely many silent steps.
On the interpretation side, \citet{yoonFormalReasoningLayered2022} make layered ITree-based interpretation modular by generically lifting interpreters through monad-transformer stacks and reusing relational reasoning principles across layers.
To handle nondeterminism, \citet{chappeChoiceTreesRepresenting2023,chappeChoiceTreesRepresenting2025a} propose \emph{Choice Trees} (CTrees), an extension of ITrees serving as a semantic domain for effectful computations with internal nondeterministic choices. Here, \emph{internal} means that the nondeterminism is \emph{not} an observable effect, in contrast to nondeterministic \coqtt{Vis} event nodes, which must remain observable. From the labeled-transition-system perspective, CTrees are equipped with rich theories of heterogeneous (bi)similarity, making them a natural domain for representing and reasoning about automata -- and hence for our \ibuchi{} specifications.
To accommodate diverse computation patterns, \citet{liProgramAdverbsTlon2022} propose compositional \emph{program adverbs} and \emph{Tl\"on embeddings}, which allow more flexible modeling of effects while retaining more information about the syntactic structure of programs.
\citet{fruminModularDenotationalSemantics2024} and \citet{stepanenkoContextDependentEffectsGuarded2025} extend the ITree-style approach from first-order effectful computations to higher-order language semantics. They use \emph{Guarded Interaction Trees} and logical relations based on Iris \cite{jungIrisGroundModular2018} to model higher-order effects -- later extended to context-dependent effects -- while preserving support for modular reasoning.

\paragraph{Modular Program Logics}
Dijkstra Monads \cite{ahmanDijkstraMonadsFree2017,jacobsDijkstraHoareMonads2015,maillardDijkstraMonadsAll2019,swamyVerifyingHigherorderPrograms2013} provide a foundation for deriving weakest-precondition transformers for arbitrary monadic computations.
Building on this methodology, \citet{silverDijkstraMonadsForever2021} propose well-behaved specification monads for ITrees whose event signature is either empty (in which case they coincide with the delay monad) or carries uninterpreted events. For the latter case, they develop the \coqtt{TraceSpec} monad based on the trace semantics of ITrees, on which we build our temporal logic.
\citet{maillardNext700Relational2019} extend the study of program logics from single effectful programs to pairs of programs, thereby yielding a powerful framework of relational program logics.
\citet{vistrupProgramLogicsCarte2025} develop \emph{Program Logics {\`a} la Carte} by decomposing language semantics and proof rules into reusable effect-specific fragments, using ITrees for semantics and Iris \cite{jungIrisGroundModular2018} for soundness, so that logics for languages with features such as shared state, concurrency, and nondeterminism can be constructed modularly rather than from scratch.

\paragraph{Temporal Program Logics}
Our work draws its main inspiration from the line of work on temporal program logics developed by \citet{songAutomatedTemporalVerification2022,songProveNFixTemporalPropertyGuided2024,songSpecifyingVerifyingFuture2026}.
In particular, \citet{songAutomatedTemporalVerification2022} propose \emph{ContEffs}, a trace-based temporal program logic for algebraic effects and handlers, whereas our work develops similar temporal reasoning over ITrees and explores temporal logic as a reusable specification component for general-purpose semantic frameworks.
Among earlier studies of trace properties,
\citet{malechaTracebasedVerificationImperative2011} focus on finite (terminating) traces, while
\citet{nakataHoareLogicCoinductive2010} focus on infinite (diverging) ones.
The semantics of the temporal formulas in \cite{songAutomatedTemporalVerification2022} is essentially \emph{coinductive}, although this is not explicitly stated or mechanized.
Our \ibuchi{} type and combinators capture this coinductive nature and can express all three variants of the star combinator, fully mechanized in \rocq.
\citet{songProveNFixTemporalPropertyGuided2024,songSpecifyingVerifyingFuture2026} introduce and develop the concept of \emph{future conditions} -- modular temporal specifications that complement pre-/post-conditions by describing expected behaviors after a function call -- enabling compositional static analysis, API-misuse detection, and proof-based repair. We also develop a generator-based, event-level encoding of future conditions, embedded in \coqtt{TraceSpec}, and demonstrate mixed specifications with state pre-/post-conditions through a case study.
\citet{wolperTemporalLogicCan1983} proves that regular expressions subsume classical LTL in expressive power. Our design follows this result: we first define elementary regular-expression combinators and then derive the LTL modalities from them, in a spirit similar to \citet{songAutomatedTemporalVerification2022,songProveNFixTemporalPropertyGuided2024,songSpecifyingVerifyingFuture2026}.


\section{Conclusion}
\label{sec:conclusion}

We have presented a compositional framework for specifying and verifying temporal properties of effectful programs whose semantics is given as ITrees. Our framework centers on \ibuchi{}, a temporal specification language shallowly embedded into CTrees, which supports sequencing, branching, conjunction, (in)finite repetition, and LTL modalities; it comes with trace- and tree-level satisfaction relations and compositional reasoning principles, and integrates into the Dijkstra Monads Forever framework, where temporal specifications compose with Hoare-style pre-/post-conditions and generator-based future conditions. The entire development is mechanized in Rocq.


\paragraph{Limitations.}
Our framework currently has two main limitations.
First, while the proposed rules support compositional reasoning for many trace- and tree-level satisfaction proofs, they do not yet form a complete high-level proof system; in some cases, users still need to fall back to low-level coinductive reasoning.
Second, our $\iter$ rules are not yet as general as desired; identifying more flexible rules for iterative programs remains future work.

\paragraph{Future Work.}
We envision three main directions:
\begin{enumerate*}[label=(\roman*)]
  \item Relating our trace-based temporal specifications to other verification frameworks -- such as separation logic (via Iris \cite{jungIrisGroundModular2018}) and higher-order semantics (via GITrees \cite{fruminModularDenotationalSemantics2024}) -- for specifying richer temporal, resource-sensitive, or higher-order properties of effectful programs.
  \item It remains unexplored how temporal properties of a program with multiple effects can be transported through partial interpretation, yielding properties of the residual semantics with fewer uninterpreted effects.
  \item Extending our trace-based LTL-style compositional temporal logic to more expressive branching-time and fixpoint logics, such as \emph{Computation Tree Logic*} (CTL*) and the \emph{modal $\mu$-calculus}~\cite{clarkeHandbookModelChecking2018a}, may further broaden its applicability to richer temporal properties.
\end{enumerate*}

\section{Data-Availability Statement}\label{sec:data-availability}

Our artifact in the \rocq{} is available at \artifacturl.

\bibliographystyle{ACM-Reference-Format}
\bibliography{body/ITrees,body/Manual}


\begin{thebibliography}{42}


\ifx \showCODEN    \undefined \def \showCODEN     #1{\unskip}     \fi
\ifx \showISBNx    \undefined \def \showISBNx     #1{\unskip}     \fi
\ifx \showISBNxiii \undefined \def \showISBNxiii  #1{\unskip}     \fi
\ifx \showISSN     \undefined \def \showISSN      #1{\unskip}     \fi
\ifx \showLCCN     \undefined \def \showLCCN      #1{\unskip}     \fi
\ifx \shownote     \undefined \def \shownote      #1{#1}          \fi
\ifx \showarticletitle \undefined \def \showarticletitle #1{#1}   \fi
\ifx \showURL      \undefined \def \showURL       {\relax}        \fi
\providecommand\bibfield[2]{#2}
\providecommand\bibinfo[2]{#2}
\providecommand\natexlab[1]{#1}
\providecommand\showeprint[2][]{arXiv:#2}

\bibitem[Ahman et~al\mbox{.}(2017)]%
        {ahmanDijkstraMonadsFree2017}
\bibfield{author}{\bibinfo{person}{Danel Ahman}, \bibinfo{person}{C{\u a}t{\u
  a}lin Hri{\c t}cu}, \bibinfo{person}{Kenji Maillard}, \bibinfo{person}{Guido
  Mart{\'i}nez}, \bibinfo{person}{Gordon Plotkin}, \bibinfo{person}{Jonathan
  Protzenko}, \bibinfo{person}{Aseem Rastogi}, {and} \bibinfo{person}{Nikhil
  Swamy}.} \bibinfo{year}{2017}\natexlab{}.
\newblock \showarticletitle{Dijkstra Monads for Free}.
\newblock \bibinfo{journal}{\emph{SIGPLAN Not.}} \bibinfo{volume}{52},
  \bibinfo{number}{1} (\bibinfo{date}{Jan.} \bibinfo{year}{2017}),
  \bibinfo{pages}{515--529}.
\newblock
\showISSN{0362-1340}
\href{https://doi.org/10.1145/3093333.3009878}{doi:\nolinkurl{10.1145/3093333.3009878}}


\bibitem[Chappe et~al\mbox{.}(2025a)]%
        {chappeChoiceTreesRepresenting2025a}
\bibfield{author}{\bibinfo{person}{Nicolas Chappe}, \bibinfo{person}{Paul He},
  \bibinfo{person}{Ludovic Henrio}, \bibinfo{person}{Eleftherios Ioannidis},
  \bibinfo{person}{Yannick Zakowski}, {and} \bibinfo{person}{Steve Zdancewic}.}
  \bibinfo{year}{2025}\natexlab{a}.
\newblock \showarticletitle{Choice Trees: {{Representing}} and Reasoning about
  Nondeterministic, Recursive, and Impure Programs in {{Rocq}}}.
\newblock \bibinfo{journal}{\emph{Journal of Functional Programming}}
  \bibinfo{volume}{35} (\bibinfo{date}{Jan.} \bibinfo{year}{2025}),
  \bibinfo{pages}{e21}.
\newblock
\showISSN{0956-7968, 1469-7653}
\href{https://doi.org/10.1017/S0956796825100105}{doi:\nolinkurl{10.1017/S0956796825100105}}


\bibitem[Chappe et~al\mbox{.}(2023)]%
        {chappeChoiceTreesRepresenting2023}
\bibfield{author}{\bibinfo{person}{Nicolas Chappe}, \bibinfo{person}{Paul He},
  \bibinfo{person}{Ludovic Henrio}, \bibinfo{person}{Yannick Zakowski}, {and}
  \bibinfo{person}{Steve Zdancewic}.} \bibinfo{year}{2023}\natexlab{}.
\newblock \showarticletitle{Choice {{Trees}}: {{Representing
  Nondeterministic}}, {{Recursive}}, and {{Impure Programs}} in {{Coq}}}.
\newblock \bibinfo{journal}{\emph{Proc. ACM Program. Lang.}}
  \bibinfo{volume}{7}, \bibinfo{number}{POPL} (\bibinfo{date}{Jan.}
  \bibinfo{year}{2023}), \bibinfo{pages}{61:1770--61:1800}.
\newblock
\href{https://doi.org/10.1145/3571254}{doi:\nolinkurl{10.1145/3571254}}


\bibitem[Chappe et~al\mbox{.}(2025b)]%
        {chappeMonadicInterpretersConcurrent2025}
\bibfield{author}{\bibinfo{person}{Nicolas Chappe}, \bibinfo{person}{Ludovic
  Henrio}, {and} \bibinfo{person}{Yannick Zakowski}.}
  \bibinfo{year}{2025}\natexlab{b}.
\newblock \showarticletitle{Monadic {{Interpreters}} for {{Concurrent Memory
  Models}}: {{Executable Semantics}} of a {{Concurrent Subset}} of {{LLVM
  IR}}}. In \bibinfo{booktitle}{\emph{Proceedings of the 14th {{ACM SIGPLAN
  International Conference}} on {{Certified Programs}} and {{Proofs}}}}
  \emph{(\bibinfo{series}{{{CPP}} '25})}. \bibinfo{publisher}{Association for
  Computing Machinery}, \bibinfo{address}{New York, NY, USA},
  \bibinfo{pages}{283--298}.
\newblock
\showISBNx{979-8-4007-1347-7}
\href{https://doi.org/10.1145/3703595.3705890}{doi:\nolinkurl{10.1145/3703595.3705890}}


\bibitem[Clarke(1997)]%
        {clarkeModelChecking1997}
\bibfield{author}{\bibinfo{person}{Edmund~M. Clarke}.}
  \bibinfo{year}{1997}\natexlab{}.
\newblock \showarticletitle{Model Checking}. In
  \bibinfo{booktitle}{\emph{Foundations of {{Software Technology}} and
  {{Theoretical Computer Science}}}},
  \bibfield{editor}{\bibinfo{person}{S.~Ramesh} {and}
  \bibinfo{person}{G.~Sivakumar}} (Eds.). \bibinfo{publisher}{Springer},
  \bibinfo{address}{Berlin, Heidelberg}, \bibinfo{pages}{54--56}.
\newblock
\showISBNx{978-3-540-69659-9}
\href{https://doi.org/10.1007/BFb0058022}{doi:\nolinkurl{10.1007/BFb0058022}}


\bibitem[Clarke et~al\mbox{.}(2018)]%
        {clarkeHandbookModelChecking2018a}
\bibfield{editor}{\bibinfo{person}{Edmund~M. Clarke},
  \bibinfo{person}{Thomas~A. Henzinger}, \bibinfo{person}{Helmut Veith}, {and}
  \bibinfo{person}{Roderick Bloem}} (Eds.). \bibinfo{year}{2018}\natexlab{}.
\newblock \bibinfo{booktitle}{\emph{Handbook of {{Model Checking}}}}.
\newblock \bibinfo{publisher}{Springer International Publishing},
  \bibinfo{address}{Cham}.
\newblock
\showISBNx{978-3-319-10574-1 978-3-319-10575-8}
\href{https://doi.org/10.1007/978-3-319-10575-8}{doi:\nolinkurl{10.1007/978-3-319-10575-8}}


\bibitem[Dijkstra(1975)]%
        {dijkstraGuardedCommandsNondeterminacy1975}
\bibfield{author}{\bibinfo{person}{Edsger~W. Dijkstra}.}
  \bibinfo{year}{1975}\natexlab{}.
\newblock \showarticletitle{Guarded Commands, Nondeterminacy and Formal
  Derivation of Programs}.
\newblock \bibinfo{journal}{\emph{Commun. ACM}} \bibinfo{volume}{18},
  \bibinfo{number}{8} (\bibinfo{date}{Aug.} \bibinfo{year}{1975}),
  \bibinfo{pages}{453--457}.
\newblock
\showISSN{0001-0782}
\href{https://doi.org/10.1145/360933.360975}{doi:\nolinkurl{10.1145/360933.360975}}


\bibitem[Frumin et~al\mbox{.}(2024)]%
        {fruminModularDenotationalSemantics2024}
\bibfield{author}{\bibinfo{person}{Dan Frumin}, \bibinfo{person}{Amin Timany},
  {and} \bibinfo{person}{Lars Birkedal}.} \bibinfo{year}{2024}\natexlab{}.
\newblock \showarticletitle{Modular {{Denotational Semantics}} for {{Effects}}
  with {{Guarded Interaction Trees}}}.
\newblock \bibinfo{journal}{\emph{Proc. ACM Program. Lang.}}
  \bibinfo{volume}{8}, \bibinfo{number}{POPL} (\bibinfo{date}{Jan.}
  \bibinfo{year}{2024}), \bibinfo{pages}{12:332--12:361}.
\newblock
\href{https://doi.org/10.1145/3632854}{doi:\nolinkurl{10.1145/3632854}}


\bibitem[Hoare(1969)]%
        {hoareAxiomaticBasisComputer1969}
\bibfield{author}{\bibinfo{person}{C.~A.~R. Hoare}.}
  \bibinfo{year}{1969}\natexlab{}.
\newblock \showarticletitle{An Axiomatic Basis for Computer Programming}.
\newblock \bibinfo{journal}{\emph{Commun. ACM}} \bibinfo{volume}{12},
  \bibinfo{number}{10} (\bibinfo{date}{Oct.} \bibinfo{year}{1969}),
  \bibinfo{pages}{576--580}.
\newblock
\showISSN{0001-0782}
\href{https://doi.org/10.1145/363235.363259}{doi:\nolinkurl{10.1145/363235.363259}}


\bibitem[Hur et~al\mbox{.}(2013)]%
        {hurPowerParameterizationCoinductive2013}
\bibfield{author}{\bibinfo{person}{Chung-Kil Hur}, \bibinfo{person}{Georg
  Neis}, \bibinfo{person}{Derek Dreyer}, {and} \bibinfo{person}{Viktor
  Vafeiadis}.} \bibinfo{year}{2013}\natexlab{}.
\newblock \showarticletitle{The Power of Parameterization in Coinductive
  Proof}. In \bibinfo{booktitle}{\emph{Proceedings of the 40th Annual {{ACM
  SIGPLAN-SIGACT}} Symposium on {{Principles}} of Programming Languages}}
  \emph{(\bibinfo{series}{{{POPL}} '13})}. \bibinfo{publisher}{Association for
  Computing Machinery}, \bibinfo{address}{New York, NY, USA},
  \bibinfo{pages}{193--206}.
\newblock
\showISBNx{978-1-4503-1832-7}
\href{https://doi.org/10.1145/2429069.2429093}{doi:\nolinkurl{10.1145/2429069.2429093}}


\bibitem[{ITrees Development Team}(2026a)]%
        {itraceDefinitionCoqdoc}
\bibfield{author}{\bibinfo{person}{{ITrees Development Team}}.}
  \bibinfo{year}{2026}\natexlab{a}.
\newblock \bibinfo{title}{{ITree.Extra.ITrace.ITraceDefinition}}.
\newblock
  \bibinfo{howpublished}{\url{https://deepspec.github.io/InteractionTrees/master/ITree.Extra.ITrace.ITraceDefinition.html}}.
\newblock
\shownote{Coqdoc documentation for the Interaction Trees library. Accessed:
  2026-07-08}.
\newblock


\bibitem[{ITrees Development Team}(2026b)]%
        {itreeInfiniteCoqdoc}
\bibfield{author}{\bibinfo{person}{{ITrees Development Team}}.}
  \bibinfo{year}{2026}\natexlab{b}.
\newblock \bibinfo{title}{{ITree.Props.Infinite}}.
\newblock
  \bibinfo{howpublished}{\url{https://deepspec.github.io/InteractionTrees/master/ITree.Props.Infinite.html}}.
\newblock
\shownote{Coqdoc documentation for the Interaction Trees library. Accessed:
  2026-07-08}.
\newblock


\bibitem[Jacobs(2015)]%
        {jacobsDijkstraHoareMonads2015}
\bibfield{author}{\bibinfo{person}{Bart Jacobs}.}
  \bibinfo{year}{2015}\natexlab{}.
\newblock \showarticletitle{Dijkstra and {{Hoare}} Monads in Monadic
  Computation}.
\newblock \bibinfo{journal}{\emph{Theoretical Computer Science}}
  \bibinfo{volume}{604} (\bibinfo{date}{Nov.} \bibinfo{year}{2015}),
  \bibinfo{pages}{30--45}.
\newblock
\showISSN{0304-3975}
\href{https://doi.org/10.1016/j.tcs.2015.03.020}{doi:\nolinkurl{10.1016/j.tcs.2015.03.020}}


\bibitem[Jung et~al\mbox{.}(2018)]%
        {jungIrisGroundModular2018}
\bibfield{author}{\bibinfo{person}{Ralf Jung}, \bibinfo{person}{Robbert
  Krebbers}, \bibinfo{person}{Jacques-Henri Jourdan}, \bibinfo{person}{Ale{\v
  s} Bizjak}, \bibinfo{person}{Lars Birkedal}, {and} \bibinfo{person}{Derek
  Dreyer}.} \bibinfo{year}{2018}\natexlab{}.
\newblock \showarticletitle{Iris from the Ground up: {{A}} Modular Foundation
  for Higher-Order Concurrent Separation Logic}.
\newblock \bibinfo{journal}{\emph{Journal of Functional Programming}}
  \bibinfo{volume}{28} (\bibinfo{date}{Jan.} \bibinfo{year}{2018}),
  \bibinfo{pages}{e20}.
\newblock
\showISSN{0956-7968, 1469-7653}
\href{https://doi.org/10.1017/S0956796818000151}{doi:\nolinkurl{10.1017/S0956796818000151}}


\bibitem[Li and Weirich(2022)]%
        {liProgramAdverbsTlon2022}
\bibfield{author}{\bibinfo{person}{Yao Li} {and} \bibinfo{person}{Stephanie
  Weirich}.} \bibinfo{year}{2022}\natexlab{}.
\newblock \showarticletitle{Program Adverbs and {{Tl\"on}} Embeddings}.
\newblock \bibinfo{journal}{\emph{Proc. ACM Program. Lang.}}
  \bibinfo{volume}{6}, \bibinfo{number}{ICFP} (\bibinfo{date}{Aug.}
  \bibinfo{year}{2022}), \bibinfo{pages}{101:312--101:342}.
\newblock
\href{https://doi.org/10.1145/3547632}{doi:\nolinkurl{10.1145/3547632}}


\bibitem[Liang et~al\mbox{.}(1995)]%
        {liangMonadTransformersModular1995}
\bibfield{author}{\bibinfo{person}{Sheng Liang}, \bibinfo{person}{Paul Hudak},
  {and} \bibinfo{person}{Mark Jones}.} \bibinfo{year}{1995}\natexlab{}.
\newblock \showarticletitle{Monad Transformers and Modular Interpreters}. In
  \bibinfo{booktitle}{\emph{Proceedings of the 22nd {{ACM SIGPLAN-SIGACT}}
  Symposium on {{Principles}} of Programming Languages}}
  \emph{(\bibinfo{series}{{{POPL}} '95})}. \bibinfo{publisher}{Association for
  Computing Machinery}, \bibinfo{address}{New York, NY, USA},
  \bibinfo{pages}{333--343}.
\newblock
\showISBNx{978-0-89791-692-9}
\href{https://doi.org/10.1145/199448.199528}{doi:\nolinkurl{10.1145/199448.199528}}


\bibitem[Maillard et~al\mbox{.}(2019a)]%
        {maillardDijkstraMonadsAll2019}
\bibfield{author}{\bibinfo{person}{Kenji Maillard}, \bibinfo{person}{Danel
  Ahman}, \bibinfo{person}{Robert Atkey}, \bibinfo{person}{Guido Mart{\'i}nez},
  \bibinfo{person}{C{\u a}t{\u a}lin Hri{\c t}cu}, \bibinfo{person}{Exequiel
  Rivas}, {and} \bibinfo{person}{{\'E}ric Tanter}.}
  \bibinfo{year}{2019}\natexlab{a}.
\newblock \showarticletitle{Dijkstra Monads for All}.
\newblock \bibinfo{journal}{\emph{Proc. ACM Program. Lang.}}
  \bibinfo{volume}{3}, \bibinfo{number}{ICFP} (\bibinfo{date}{July}
  \bibinfo{year}{2019}), \bibinfo{pages}{104:1--104:29}.
\newblock
\href{https://doi.org/10.1145/3341708}{doi:\nolinkurl{10.1145/3341708}}


\bibitem[Maillard et~al\mbox{.}(2019b)]%
        {maillardNext700Relational2019}
\bibfield{author}{\bibinfo{person}{Kenji Maillard}, \bibinfo{person}{C{\u
  a}t{\u a}lin Hri{\c t}cu}, \bibinfo{person}{Exequiel Rivas}, {and}
  \bibinfo{person}{Antoine Van~Muylder}.} \bibinfo{year}{2019}\natexlab{b}.
\newblock \showarticletitle{The next 700 Relational Program Logics}.
\newblock \bibinfo{journal}{\emph{Proc. ACM Program. Lang.}}
  \bibinfo{volume}{4}, \bibinfo{number}{POPL} (\bibinfo{date}{Dec.}
  \bibinfo{year}{2019}), \bibinfo{pages}{4:1--4:33}.
\newblock
\href{https://doi.org/10.1145/3371072}{doi:\nolinkurl{10.1145/3371072}}


\bibitem[Malecha et~al\mbox{.}(2011)]%
        {malechaTracebasedVerificationImperative2011}
\bibfield{author}{\bibinfo{person}{Gregory Malecha}, \bibinfo{person}{Greg
  Morrisett}, {and} \bibinfo{person}{Ryan Wisnesky}.}
  \bibinfo{year}{2011}\natexlab{}.
\newblock \showarticletitle{Trace-Based Verification of Imperative Programs
  with {{I}}/{{O}}}.
\newblock \bibinfo{journal}{\emph{Journal of Symbolic Computation}}
  \bibinfo{volume}{46}, \bibinfo{number}{2} (\bibinfo{year}{2011}),
  \bibinfo{pages}{95--118}.
\newblock


\bibitem[Marriott et~al\mbox{.}(2003)]%
        {marriottResourceUsageVerification2003}
\bibfield{author}{\bibinfo{person}{Kim Marriott}, \bibinfo{person}{Peter~J.
  Stuckey}, {and} \bibinfo{person}{Martin Sulzmann}.}
  \bibinfo{year}{2003}\natexlab{}.
\newblock \showarticletitle{Resource {{Usage Verification}}}. In
  \bibinfo{booktitle}{\emph{Programming {{Languages}} and {{Systems}}}},
  \bibfield{editor}{\bibinfo{person}{Atsushi Ohori}} (Ed.).
  \bibinfo{publisher}{Springer}, \bibinfo{address}{Berlin, Heidelberg},
  \bibinfo{pages}{212--229}.
\newblock
\showISBNx{978-3-540-40018-9}
\href{https://doi.org/10.1007/978-3-540-40018-9_15}{doi:\nolinkurl{10.1007/978-3-540-40018-9_15}}


\bibitem[Moggi(1989)]%
        {moggiComputationalLambdacalculusMonads1989}
\bibfield{author}{\bibinfo{person}{E. Moggi}.} \bibinfo{year}{1989}\natexlab{}.
\newblock \showarticletitle{Computational Lambda-Calculus and Monads}. In
  \bibinfo{booktitle}{\emph{Proceedings of the {{Fourth Annual Symposium}} on
  {{Logic}} in Computer Science}}. \bibinfo{publisher}{IEEE Press},
  \bibinfo{address}{Pacific Grove, California, USA}, \bibinfo{pages}{14--23}.
\newblock
\showISBNx{978-0-8186-1954-0}


\bibitem[Moggi(1991)]%
        {moggiNotionsComputationMonads1991}
\bibfield{author}{\bibinfo{person}{Eugenio Moggi}.}
  \bibinfo{year}{1991}\natexlab{}.
\newblock \showarticletitle{Notions of Computation and Monads}.
\newblock \bibinfo{journal}{\emph{Information and Computation}}
  \bibinfo{volume}{93}, \bibinfo{number}{1} (\bibinfo{date}{July}
  \bibinfo{year}{1991}), \bibinfo{pages}{55--92}.
\newblock
\showISSN{08905401}
\href{https://doi.org/10.1016/0890-5401(91)90052-4}{doi:\nolinkurl{10.1016/0890-5401(91)90052-4}}


\bibitem[Nakata and Uustalu(2010)]%
        {nakataHoareLogicCoinductive2010}
\bibfield{author}{\bibinfo{person}{Keiko Nakata} {and} \bibinfo{person}{Tarmo
  Uustalu}.} \bibinfo{year}{2010}\natexlab{}.
\newblock \showarticletitle{A {{Hoare Logic}} for the {{Coinductive Trace-Based
  Big-Step Semantics}} of {{While}}}. In \bibinfo{booktitle}{\emph{Programming
  {{Languages}} and {{Systems}}}}, \bibfield{editor}{\bibinfo{person}{Andrew~D.
  Gordon}} (Ed.). \bibinfo{publisher}{Springer}, \bibinfo{address}{Berlin,
  Heidelberg}, \bibinfo{pages}{488--506}.
\newblock
\showISBNx{978-3-642-11957-6}
\href{https://doi.org/10.1007/978-3-642-11957-6_26}{doi:\nolinkurl{10.1007/978-3-642-11957-6_26}}


\bibitem[Pierce et~al\mbox{.}(2025)]%
        {Pierce:SF1}
\bibfield{author}{\bibinfo{person}{Benjamin~C. Pierce},
  \bibinfo{person}{Arthur~Azevedo {de Amorim}}, \bibinfo{person}{Chris
  Casinghino}, \bibinfo{person}{Marco Gaboardi}, \bibinfo{person}{Michael
  Greenberg}, \bibinfo{person}{C{\u a}t{\u a}lin Hri{\c t}cu},
  \bibinfo{person}{Vilhelm Sj{\"o}berg}, {and} \bibinfo{person}{Brent Yorgey}.}
  \bibinfo{year}{2025}\natexlab{}.
\newblock \bibinfo{booktitle}{\emph{Logical Foundations}}.
  \bibinfo{series}{Software Foundations}, Vol.~\bibinfo{volume}{1}.
\newblock \bibinfo{publisher}{Electronic textbook}.
\newblock


\bibitem[Pous(2016)]%
        {damienpousCoinductionAllWay2016}
\bibfield{author}{\bibinfo{person}{Damien Pous}.}
  \bibinfo{year}{2016}\natexlab{}.
\newblock \showarticletitle{Coinduction All the Way Up}. In
  \bibinfo{booktitle}{\emph{Proceedings of the 31st {{Annual ACM}}/{{IEEE
  Symposium}} on {{Logic}} in {{Computer Science}}}}
  \emph{(\bibinfo{series}{{{LICS}} '16})}. \bibinfo{publisher}{Association for
  Computing Machinery}, \bibinfo{address}{New York, NY, USA},
  \bibinfo{pages}{307--316}.
\newblock
\showISBNx{978-1-4503-4391-6}
\href{https://doi.org/10.1145/2933575.2934564}{doi:\nolinkurl{10.1145/2933575.2934564}}


\bibitem[{Rocq Development Team}(2026a)]%
        {rocq-prover-refman-9-2}
\bibfield{author}{\bibinfo{person}{{Rocq Development Team}}.}
  \bibinfo{year}{2026}\natexlab{a}.
\newblock \bibinfo{booktitle}{\emph{{The Rocq Prover Reference Manual}}}.
\newblock The Rocq Prover.
\newblock
\shownote{Version 9.2.0}.
\newblock
\urldef\tempurl%
\url{https://rocq-prover.org/doc/V9.2.0/refman/index.html}
\showURL{%
\tempurl}


\bibitem[{Rocq Development Team}(2026b)]%
        {rocq-prover-coinductive-9-2}
\bibfield{author}{\bibinfo{person}{{Rocq Development Team}}.}
  \bibinfo{year}{2026}\natexlab{b}.
\newblock \bibinfo{booktitle}{\emph{{The Rocq Prover Reference Manual. Core
  Language. Coinductive Types and Corecursive Functions.}}}
\newblock The Rocq Prover.
\newblock
\shownote{Version 9.2.0}.
\newblock
\urldef\tempurl%
\url{https://rocq-prover.org/doc/V9.2.0/refman/language/core/coinductive.html}
\showURL{%
\tempurl}


\bibitem[Silver and Zdancewic(2021)]%
        {silverDijkstraMonadsForever2021}
\bibfield{author}{\bibinfo{person}{Lucas Silver} {and} \bibinfo{person}{Steve
  Zdancewic}.} \bibinfo{year}{2021}\natexlab{}.
\newblock \showarticletitle{Dijkstra Monads Forever: Termination-Sensitive
  Specifications for Interaction Trees}.
\newblock \bibinfo{journal}{\emph{Proc. ACM Program. Lang.}}
  \bibinfo{volume}{5}, \bibinfo{number}{POPL} (\bibinfo{date}{Jan.}
  \bibinfo{year}{2021}), \bibinfo{pages}{26:1--26:28}.
\newblock
\href{https://doi.org/10.1145/3434307}{doi:\nolinkurl{10.1145/3434307}}


\bibitem[Song et~al\mbox{.}(2022)]%
        {songAutomatedTemporalVerification2022}
\bibfield{author}{\bibinfo{person}{Yahui Song}, \bibinfo{person}{Darius Foo},
  {and} \bibinfo{person}{Wei-Ngan Chin}.} \bibinfo{year}{2022}\natexlab{}.
\newblock \showarticletitle{Automated {{Temporal Verification}} for~{{Algebraic
  Effects}}}. In \bibinfo{booktitle}{\emph{Programming {{Languages}} and
  {{Systems}}}}, \bibfield{editor}{\bibinfo{person}{Ilya Sergey}} (Ed.).
  \bibinfo{publisher}{Springer Nature Switzerland}, \bibinfo{address}{Cham},
  \bibinfo{pages}{88--109}.
\newblock
\showISBNx{978-3-031-21037-2}
\href{https://doi.org/10.1007/978-3-031-21037-2_5}{doi:\nolinkurl{10.1007/978-3-031-21037-2_5}}


\bibitem[Song et~al\mbox{.}(2026)]%
        {songSpecifyingVerifyingFuture2026}
\bibfield{author}{\bibinfo{person}{Yahui Song}, \bibinfo{person}{Darius Foo},
  {and} \bibinfo{person}{Wei-Ngan Chin}.} \bibinfo{year}{2026}\natexlab{}.
\newblock \showarticletitle{Specifying and~{{Verifying Future Conditions}}}. In
  \bibinfo{booktitle}{\emph{Static {{Analysis}}}},
  \bibfield{editor}{\bibinfo{person}{Hakjoo Oh} {and} \bibinfo{person}{Yulei
  Sui}} (Eds.). \bibinfo{publisher}{Springer Nature Switzerland},
  \bibinfo{address}{Cham}, \bibinfo{pages}{90--112}.
\newblock
\showISBNx{978-3-032-07106-4}
\href{https://doi.org/10.1007/978-3-032-07106-4_5}{doi:\nolinkurl{10.1007/978-3-032-07106-4_5}}


\bibitem[Song et~al\mbox{.}(2024)]%
        {songProveNFixTemporalPropertyGuided2024}
\bibfield{author}{\bibinfo{person}{Yahui Song}, \bibinfo{person}{Xiang Gao},
  \bibinfo{person}{Wenhua Li}, \bibinfo{person}{Wei-Ngan Chin}, {and}
  \bibinfo{person}{Abhik Roychoudhury}.} \bibinfo{year}{2024}\natexlab{}.
\newblock \showarticletitle{{{ProveNFix}}: {{Temporal Property-Guided Program
  Repair}}}.
\newblock \bibinfo{journal}{\emph{Proc. ACM Softw. Eng.}} \bibinfo{volume}{1},
  \bibinfo{number}{FSE} (\bibinfo{date}{July} \bibinfo{year}{2024}),
  \bibinfo{pages}{11:226--11:248}.
\newblock
\href{https://doi.org/10.1145/3643737}{doi:\nolinkurl{10.1145/3643737}}


\bibitem[Stepanenko et~al\mbox{.}(2025)]%
        {stepanenkoContextDependentEffectsGuarded2025}
\bibfield{author}{\bibinfo{person}{Sergei Stepanenko}, \bibinfo{person}{Emma
  Nardino}, \bibinfo{person}{Dan Frumin}, \bibinfo{person}{Amin Timany}, {and}
  \bibinfo{person}{Lars Birkedal}.} \bibinfo{year}{2025}\natexlab{}.
\newblock \showarticletitle{Context-{{Dependent Effects}} in {{Guarded
  Interaction Trees}}}. In \bibinfo{booktitle}{\emph{Programming {{Languages}}
  and {{Systems}}}}, \bibfield{editor}{\bibinfo{person}{Viktor Vafeiadis}}
  (Ed.). \bibinfo{publisher}{Springer Nature Switzerland},
  \bibinfo{address}{Cham}, \bibinfo{pages}{286--313}.
\newblock
\showISBNx{978-3-031-91121-7}
\href{https://doi.org/10.1007/978-3-031-91121-7_12}{doi:\nolinkurl{10.1007/978-3-031-91121-7_12}}


\bibitem[Swamy et~al\mbox{.}(2013)]%
        {swamyVerifyingHigherorderPrograms2013}
\bibfield{author}{\bibinfo{person}{Nikhil Swamy}, \bibinfo{person}{Joel
  Weinberger}, \bibinfo{person}{Cole Schlesinger}, \bibinfo{person}{Juan Chen},
  {and} \bibinfo{person}{Benjamin Livshits}.} \bibinfo{year}{2013}\natexlab{}.
\newblock \showarticletitle{Verifying Higher-Order Programs with the {D}ijkstra
  Monad}. In \bibinfo{booktitle}{\emph{Proceedings of the 34th {{ACM SIGPLAN
  Conference}} on {{Programming Language Design}} and {{Implementation}}}}
  \emph{(\bibinfo{series}{{{PLDI}} '13})}. \bibinfo{publisher}{Association for
  Computing Machinery}, \bibinfo{address}{New York, NY, USA},
  \bibinfo{pages}{387--398}.
\newblock
\showISBNx{978-1-4503-2014-6}
\href{https://doi.org/10.1145/2491956.2491978}{doi:\nolinkurl{10.1145/2491956.2491978}}


\bibitem[Swierstra(2008)]%
        {swierstraDataTypesCarte2008}
\bibfield{author}{\bibinfo{person}{Wouter Swierstra}.}
  \bibinfo{year}{2008}\natexlab{}.
\newblock \showarticletitle{Data Types \`a La Carte}.
\newblock \bibinfo{journal}{\emph{J. Funct. Prog.}} \bibinfo{volume}{18},
  \bibinfo{number}{04} (\bibinfo{date}{July} \bibinfo{year}{2008}).
\newblock
\showISSN{0956-7968, 1469-7653}
\href{https://doi.org/10.1017/S0956796808006758}{doi:\nolinkurl{10.1017/S0956796808006758}}


\bibitem[{Tom Schrijvers} et~al\mbox{.}(2019)]%
        {tomschrijversMonadTransformersModular2019}
\bibfield{author}{\bibinfo{person}{{Tom Schrijvers}}, \bibinfo{person}{Maciej
  Pir{\'o}g}, \bibinfo{person}{Nicolas Wu}, {and} \bibinfo{person}{Mauro
  Jaskelioff}.} \bibinfo{year}{2019}\natexlab{}.
\newblock \showarticletitle{Monad Transformers and Modular Algebraic Effects:
  What Binds Them Together}. In \bibinfo{booktitle}{\emph{Proceedings of the
  12th {{ACM SIGPLAN International Symposium}} on {{Haskell}}}}
  \emph{(\bibinfo{series}{Haskell 2019})}. \bibinfo{publisher}{Association for
  Computing Machinery}, \bibinfo{address}{New York, NY, USA},
  \bibinfo{pages}{98--113}.
\newblock
\showISBNx{978-1-4503-6813-1}
\href{https://doi.org/10.1145/3331545.3342595}{doi:\nolinkurl{10.1145/3331545.3342595}}


\bibitem[Vistrup et~al\mbox{.}(2025)]%
        {vistrupProgramLogicsCarte2025}
\bibfield{author}{\bibinfo{person}{Max Vistrup}, \bibinfo{person}{Michael
  Sammler}, {and} \bibinfo{person}{Ralf Jung}.}
  \bibinfo{year}{2025}\natexlab{}.
\newblock \showarticletitle{Program Logics {\`{a}} la {C}arte}.
\newblock \bibinfo{journal}{\emph{Proc. {ACM} Program. Lang.}}
  \bibinfo{volume}{9}, \bibinfo{number}{{POPL}} (\bibinfo{year}{2025}),
  \bibinfo{pages}{300--331}.
\newblock
\href{https://doi.org/10.1145/3704847}{doi:\nolinkurl{10.1145/3704847}}


\bibitem[Wadler(1992)]%
        {wadlerComprehendingMonads1992}
\bibfield{author}{\bibinfo{person}{Philip Wadler}.}
  \bibinfo{year}{1992}\natexlab{}.
\newblock \showarticletitle{Comprehending Monads}.
\newblock \bibinfo{journal}{\emph{Mathematical Structures in Computer Science}}
  \bibinfo{volume}{2}, \bibinfo{number}{4} (\bibinfo{date}{Dec.}
  \bibinfo{year}{1992}), \bibinfo{pages}{461--493}.
\newblock
\showISSN{1469-8072, 0960-1295}
\href{https://doi.org/10.1017/S0960129500001560}{doi:\nolinkurl{10.1017/S0960129500001560}}


\bibitem[Wolper(1983)]%
        {wolperTemporalLogicCan1983}
\bibfield{author}{\bibinfo{person}{Pierre Wolper}.}
  \bibinfo{year}{1983}\natexlab{}.
\newblock \showarticletitle{Temporal Logic Can Be More Expressive}.
\newblock \bibinfo{journal}{\emph{Information and Control}}
  \bibinfo{volume}{56}, \bibinfo{number}{1} (\bibinfo{date}{Jan.}
  \bibinfo{year}{1983}), \bibinfo{pages}{72--99}.
\newblock
\showISSN{0019-9958}
\href{https://doi.org/10.1016/S0019-9958(83)80051-5}{doi:\nolinkurl{10.1016/S0019-9958(83)80051-5}}


\bibitem[Xia et~al\mbox{.}(2019)]%
        {xiaInteractionTreesRepresenting2019}
\bibfield{author}{\bibinfo{person}{Li-yao Xia}, \bibinfo{person}{Yannick
  Zakowski}, \bibinfo{person}{Paul He}, \bibinfo{person}{Chung-Kil Hur},
  \bibinfo{person}{Gregory Malecha}, \bibinfo{person}{Benjamin~C. Pierce},
  {and} \bibinfo{person}{Steve Zdancewic}.} \bibinfo{year}{2019}\natexlab{}.
\newblock \showarticletitle{Interaction Trees: Representing Recursive and
  Impure Programs in {{Coq}}}.
\newblock \bibinfo{journal}{\emph{Proc. ACM Program. Lang.}}
  \bibinfo{volume}{4}, \bibinfo{number}{POPL} (\bibinfo{date}{Dec.}
  \bibinfo{year}{2019}), \bibinfo{pages}{51:1--51:32}.
\newblock
\href{https://doi.org/10.1145/3371119}{doi:\nolinkurl{10.1145/3371119}}


\bibitem[Yoon et~al\mbox{.}(2022)]%
        {yoonFormalReasoningLayered2022}
\bibfield{author}{\bibinfo{person}{Irene Yoon}, \bibinfo{person}{Yannick
  Zakowski}, {and} \bibinfo{person}{Steve Zdancewic}.}
  \bibinfo{year}{2022}\natexlab{}.
\newblock \showarticletitle{Formal Reasoning about Layered Monadic
  Interpreters}.
\newblock \bibinfo{journal}{\emph{Proc. ACM Program. Lang.}}
  \bibinfo{volume}{6}, \bibinfo{number}{ICFP} (\bibinfo{date}{Aug.}
  \bibinfo{year}{2022}), \bibinfo{pages}{99:254--99:282}.
\newblock
\href{https://doi.org/10.1145/3547630}{doi:\nolinkurl{10.1145/3547630}}


\bibitem[Zakowski et~al\mbox{.}(2021)]%
        {zakowskiModularCompositionalExecutable2021}
\bibfield{author}{\bibinfo{person}{Yannick Zakowski}, \bibinfo{person}{Calvin
  Beck}, \bibinfo{person}{Irene Yoon}, \bibinfo{person}{Ilia Zaichuk},
  \bibinfo{person}{Vadim Zaliva}, {and} \bibinfo{person}{Steve Zdancewic}.}
  \bibinfo{year}{2021}\natexlab{}.
\newblock \showarticletitle{Modular, Compositional, and Executable Formal
  Semantics for {{LLVM IR}}}.
\newblock \bibinfo{journal}{\emph{Proc. ACM Program. Lang.}}
  \bibinfo{volume}{5}, \bibinfo{number}{ICFP} (\bibinfo{date}{Aug.}
  \bibinfo{year}{2021}), \bibinfo{pages}{67:1--67:30}.
\newblock
\href{https://doi.org/10.1145/3473572}{doi:\nolinkurl{10.1145/3473572}}


\bibitem[Zhang et~al\mbox{.}(2021)]%
        {zhangVerifyingHTTPKeyValue2021}
\bibfield{author}{\bibinfo{person}{Hengchu Zhang}, \bibinfo{person}{Wolf
  Honor{\'e}}, \bibinfo{person}{Nicolas Koh}, \bibinfo{person}{Yao Li},
  \bibinfo{person}{Yishuai Li}, \bibinfo{person}{Li-Yao Xia},
  \bibinfo{person}{Lennart Beringer}, \bibinfo{person}{William Mansky},
  \bibinfo{person}{Benjamin Pierce}, {and} \bibinfo{person}{Steve Zdancewic}.}
  \bibinfo{year}{2021}\natexlab{}.
\newblock \showarticletitle{Verifying an {{HTTP Key-Value Server}} with
  {{Interaction Trees}} and {{VST}}}. In \bibinfo{booktitle}{\emph{12th
  {{International Conference}} on {{Interactive Theorem Proving}} ({{ITP}}
  2021)}}. \bibinfo{publisher}{Schloss Dagstuhl -- Leibniz-Zentrum f\"ur
  Informatik}, \bibinfo{pages}{32:1--32:19}.
\newblock
\href{https://doi.org/10.4230/LIPIcs.ITP.2021.32}{doi:\nolinkurl{10.4230/LIPIcs.ITP.2021.32}}


\end{thebibliography}

\newpage
\appendix
\section*{Appendix}\label{sec:appendix}
\section{Omitted Definitions of \ibuchi{} Combinators}
\label{sec:omitted-defs}

\subsection{\ibuchi{} Negation}

The $\neg$ combinator is defined coinductively with the negation of the \coqtt{Char} type, than lifted to \ibuchi{}:

\begin{minted}[fontsize=\footnotesize,xleftmargin=2em]{coq}
Fixpoint ChrNeg {E} (c : Char E) : Char E
  := match c with
  | ChrPrd p => ChrPrd (λ R e a, ¬ p R e a)
  | ChrWld => ChrNon
  | ChrNon => ChrWld
  | ChrAnd c1 c2 => ChrOr (ChrNeg c1) (ChrNeg c2)
  | ChrOr c1 c2 => ChrAnd (ChrNeg c1) (ChrNeg c2)
  end.

CoFixpoint neg {E R} (r : iBuchi E R) : iBuchi E R
  := match observe r with
  | RetF r => Ret r
  | StepF t => Step (neg t)
  | VisF RxEme k => eme
  | VisF (RxChr c) k => Vis (RxChr (ChrNeg c)) k
  | StuckF => Stuck
  | GuardF t => Guard (neg t)
  | BrF c k => Br c (λ b, neg (k b))
  end.
Notation "'¬'  r" := (neg r) (at level 75).
\end{minted}

Note that this negation combinator is a \emph{\coqtt{Char}-wise} operation, thus not be confused with that in traditional LTL.

\subsection{\ibuchi{} Product}

The $\prd$ combinator is defined coinductively with the \coqtt{ChrAnd} constructor of the \coqtt{Char} type:

\begin{minted}[fontsize=\footnotesize,xleftmargin=2em]{coq}
CoFixpoint prd {E} (r s : iBuchi E unit) : iBuchi E unit
  := match observe r, observe s with
    | RetF _, RetF _ => Ret tt
    | RetF _, StuckF => Stuck
    | RetF _, StepF t => Step (prd r t)
    | RetF _, GuardF t => Guard (prd r t)
    | RetF _, BrF c k => Br c (λ i, prd r (k i))
    | RetF _, VisF _ _ => Stuck
    | StuckF, _ => Stuck
    | StepF t, _ => Step (prd t s)
    | GuardF t, _ => Guard (prd t s)
    | BrF c k, _ => Br c (λ i, prd (k i) s)
    | VisF _ _, RetF _ => Stuck
    | VisF _ _, StuckF => Stuck
    | VisF _ _, StepF t => Step (prd r t)
    | VisF _ _, GuardF t => Guard (prd r t)
    | VisF _ _, BrF c k => Br c (λ i, prd r (k i))
    | VisF RxEme _, VisF RxEme _ => eme
    | VisF RxEme _, _ => Stuck
    | _, VisF RxEme _ => Stuck
    | VisF (RxChr c1) k1, VisF (RxChr c2) k2 =>
        Vis (RxChr (ChrAnd c1 c2)) (λ i, prd (k1 i) (k2 i))
    end.
Infix "&" := prd (at level 69, left associativity).
\end{minted}

\section{Mechanized Lemmas in the Proofs of \textsc{TrSat-Bind} and \textsc{TrSat-Iter-Nonempty}}\label{sec:appx-lemma-in-proofs}

We list the statements of the mechanized lemmas used in our proofs, while leave the proofs of them in the ITree trace library and our Rocq implementation.

\subsection{Lemmas from the ITrees Trace Library}

\begin{lemma}[\coqtt{decompose_trace_refine_bind}]\hspace{2em}
\begin{minted}[fontsize=\footnotesize,xleftmargin=2em]{coq}
Lemma decompose_trace_refine_bind
  : ∀ (E : Type → Type) (R S : Type)
        (b : itrace E S) (t : itree E R) (f : R → itree E S),
    b ⊑ t >>= f → ∃ b', ∃ g', (ITree.bind b' g' ≈ b) ∧ b' ⊑ t.
\end{minted}
\end{lemma}

\begin{lemma}[\coqtt{trace_refine_bind_cont_inv}]\hspace{2em}
\begin{minted}[fontsize=\footnotesize,xleftmargin=2em]{coq}
Lemma trace_refine_bind_cont_inv
  : ∀ (E : Type → Type) (R S : Type)
        (b : itrace E R) (m : itree E R) (g : R → itrace E S)
        (f : R → itree E S) (r : R),
    may_converge r b → b ⊑ m → ITree.bind b g ⊑ ITree.bind m f → g r ⊑ f r.
\end{minted}
\end{lemma}

\subsection{Our Lemmas}

\begin{lemma}[\coqtt{sat_Ret_prefix}]\hspace{2em}
\begin{minted}[fontsize=\footnotesize,xleftmargin=2em]{coq}
Lemma sat_Ret_prefix {E A X}
    (t : ctrace E X) (r : iBuchi E unit) (a : A)
  : (CTree.bind t (λ _, cRet a)) ⊨ r
  → t ⊨ r.
\end{minted}
\end{lemma}

\begin{lemma}[\coqtt{trans_embed_val_inf_false}]\hspace{2em}
\begin{minted}[fontsize=\footnotesize,xleftmargin=2em]{coq}
Lemma trans_embed_val_inf_false {E A B}
    (t' : itree E A) (t u : ctree E B0 A) (b : B)
  : all_infinite t'
  → t (≅eq) embed t'
  → trans (val b) t u
  → False.
\end{minted}
\end{lemma}

\begin{lemma}[\coqtt{trans_embed_obs_inf_inf}]\hspace{2em}
\begin{minted}[fontsize=\footnotesize,xleftmargin=2em]{coq}
Lemma trans_embed_obs_inf_inf {E A B}
    (e : E A) (a : A) (t' : itree E B) (t u : ctree E B0 B)
  : all_infinite t'
  → t (≅eq) embed t'
  → trans (obs e a) t u
  → ∃ u' : itree E B,
    u (≅eq) Guard (embed u') ∧ all_infinite u'.
\end{minted}
\end{lemma}

\begin{lemma}[\coqtt{sat_bind_divergent}]\hspace{2em}
\begin{minted}[fontsize=\footnotesize,xleftmargin=2em]{coq}
Lemma sat_bind_divergent {E A B}
    (tc : itrace E A) (ktc : A → itrace E B) (r s : iBuchi E unit)
  : all_infinite tc
  → embed tc ⊨ r
  → embed (ITree.bind tc ktc) ⊨ (r ++ s).
\end{minted}
\end{lemma}

\begin{lemma}[\coqtt{SatL_is_update_val_rel}]\hspace{2em}
\begin{minted}[fontsize=\footnotesize,xleftmargin=2em]{coq}
Lemma SatL_is_update_val_rel {E R S}
  : @is_update_val_rel (EvAns E) (EvRx E) R S
    (SatL (E:=E)) (λ _ _, True) (SatL (E:=E)).
\end{minted}
\end{lemma}









\end{document}